\documentclass[11pt,a4paper]{article}

\usepackage{graphicx}
\usepackage{xcolor}
\usepackage{amsmath}
\usepackage{amssymb}
\usepackage{tikz}
\usepackage{ifthen}
\usepackage{natbib}

\begin{document}

\title{Influence of the approach boundary layer on the flow over an axisymmetric hill at a moderate Reynolds number}

\date{}

\author{M. Garc\'{\i}a-Villalba$^{\rm a}$ , J.~G.~Wissink$^{\rm b}$ and W. Rodi$^{\rm a}$ \\
$^{\rm a}${\em{Karlsruhe Institute of Technology,Germany}}\\
$^{\rm b}${\em{Brunel University, United Kingdom}}
} 


\maketitle

\begin{abstract}
Large Eddy Simulations of a flow at a moderate Reynolds number over and around 
a three-dimensional hill have been performed. The main aim of the simulations 
was to study the effects of various inflow conditions (boundary layer thickness
and laminar versus turbulent boundary layers) on the flow behind the hill.
The main features of the flow behind the hill are similar in all simulations,
however various differences are observed.
The topology of the streamlines (friction lines) on the surface adjacent to 
the lower wall was found to be independent of the inflow conditions prescribed 
and comprised four saddle points and four nodes (of which two are foci).
In all simulations a variety of vortical structures could be observed, ranging
from a horseshoe vortex - that was formed at the foot of the hill - to a train of
large hairpin vortices in the wake of the hill.
In the simulation with a thick incoming laminar boundary layer also secondary
vortical structures (i.e. hairpin vortices) were observed to be formed at 
either side of the hill, superposed on the legs of the horseshoe vortex.  
Sufficiently far downstream of the hill, at the symmetry plane 
the mean velocity and the 
rms of the velocity fluctuations were found to become quasi-independent of the inflow
conditions, while towards the sides the influence of the hill decreases
and the velocity profiles recover the values prevailing at the inflow.    
\end{abstract}

\unitlength 1mm

\section{Introduction}

Flow over obstacles occurs in many engineering applications. 
From a fundamental point of view, the flow over obstacles
features a variety of phenomena and is particularly complex:
it is three-dimensional (also in the mean), highly unsteady,
involves separation and  reattachment (possibly at several locations)
and contains several interacting vortex systems. 
Typical examples are flow over axisymmetric obstacles \cite{hunt:80},
wall-mounted prismatic obstacles 
\cite{martinuzzi:93} or finite-height circular cylinders \cite{palau:09}.
In the case of obstacles without sharp edges, the separation location
is not fixed by the geometry and is highly dependent on the incoming 
flow characteristics, like Reynolds number or boundary layer thickness.
Therefore, it is very challenging to compute accurately this kind of flow.

Recently, a series of experiments performed by Simpson and co-workers 
\cite{simpson:02,byun:04,ma:05,byun:06} renewed the interest 
in analysing and predicting three-dimensional separation. 
The configuration they considered was an axisymmetric three-dimensional
hill subjected to a turbulent boundary layer.
The Reynolds number of the flow based on the free-stream velocity and the height of the hill was relatively high ($Re=130 000$). Since then
a number of researchers have tried to reproduce the experimental results
using various computational  techniques: 
RANS \cite{wang:04,persson:06}, hybrid RANS-LES \cite{tessicini:07}
and pure LES \cite{persson:06,patel:07,krajnovic:08,gv:09}.
It was shown that the RANS predictions were generally poor, while both 
the hybrid techniques and the pure LES provided promising results 
although not completely satisfactory. In the simulations reported in 
\cite{gv:09}, it was observed that, in spite
of the high resolution employed, the presence of a 
thin recirculation region made the flow very sensitive to the grid resolution.

Apart from the grid resolution, one of the most important differences
between all computational studies was the modelling of the incoming flow. 
In the experiment the hill was subjected to a turbulent boundary layer
whose thickness was half of the hill height. 
Because the incoming boundary layer
was turbulent, the specification of the inlet 
conditions had to be done in an unsteady 
manner. 
For instance, in \cite{gv:09} a precursor simulation was used, while
in \cite{patel:07} a boundary layer profile plus random noise was employed. 
Because of the high Reynolds number of the
flow, it is computationally very expensive 
to perform parametric studies and, hence,
it is very difficult to assess the impact
on the results of the modelling of the incoming flow. 

In the present paper, we report simulations of flow over the same hill as 
considered in the previous studies, 
but at a significantly lower Reynolds number. The aim is to study
the influence of the flow characteristics of various approach flows: 
two simulations with incoming laminar boundary layers of different thicknesses
and one simulation with an incoming turbulent boundary layer.

\section{Numerical model}

The LES
were performed with the in-house code LESOCC2
(Large Eddy Simulation On Curvilinear Coordinates).
The code has been developed at the Institute
for Hydromechanics. It is the successor of the code LESOCC developed by
Breuer and Rodi \cite{breuer:96} and is described 
in \cite{hinterberger:04}. The code solves the Navier-Stokes equations
on body-fitted, curvilinear grids using a
cell-centered Finite Volume method with collocated storage for the
cartesian velocity components and the pressure. 
Second order central differences
are employed for the convection as well as for the diffusive terms. 
The time integration is performed with a predictor-corrector scheme, where
the explicit predictor step for the momentum equations is a low-storage 3-step
Runge-Kutta method. The corrector step covers the implicit solution of the Poisson
equation for the pressure correction (SIMPLE). The scheme is of second order accuracy in time
because the Poisson equation for the pressure correction is not solved during the sub-steps
of the Runge-Kutta algorithm in order to save CPU-time.
The Rhie and Chow momentum
interpolation \cite{rhie:83} is applied to avoid pressure-velocity
decoupling. The Poisson equation for the pressure-increment is
solved iteratively by means of the 'strongly implicit procedure'
\cite{stone:68}. Parallelization is implemented via domain
decomposition, and explicit message passing is used with two halo
cells along the inter-domain boundaries for intermediate storage.

The configuration mentioned above consists of the flow over and around an 
axisymmetric hill of height H and base-to-height
ratio of 4.
The hill shape is described by:
\begin{equation}
\frac{y(r)}{H}=-\frac{1}{6.04844}\left [ 
J_0(\Lambda) I_0 \left ( \Lambda \frac{r}{2H}\right ) 
- I_0(\Lambda) J_0 \left ( \Lambda \frac{r}{2H}\right )  \right ],
\end{equation}
where $\Lambda=3.1926$, $J_0$ is the Bessel function of the first kind and $I_0$ is the modified 
Bessel function of the first kind \cite{simpson:02}.
The approach-flow boundary-layer 
has a thickness $\delta$ which varies depending on the simulation (see Table \ref{tab:1}). 
The Reynolds number of the flow based on 
the free-stream velocity $U_{ref}$ and the hill height $H$ is $Re=6650$.

The size of the domain
is $22 H \times 6 H \times 12 H$ in streamwise, wall-normal and spanwise directions, respectively.
The grid consists of $504 \times 256 \times 400$ cells in these directions.
The choice of the computational mesh is based on the experience gained
in performing an LES of a similar flow problem at a significantly higher $Re$ 
\cite{gv:09}, with results compared to experimental data. 
 In order 
to minimize numerical errors, 
the grid is quasi-orthogonal close to the hill's surface
and the grid points are concentrated in the boundary layer. 

As in \cite{gv:09}, the dynamic Smagorinsky subgrid-scale model, 
first proposed by Germano et al. \cite{germano:91}, has been used in the simulations.
The model parameter is determined using an explicit box filter of width equal
to twice the mesh size and smoothed by temporal under-relaxation \cite{breuer:96}.
The impact of the subgrid-scale model on the results 
is likely to be small. For instance,
the maximum values of the ratio of the time-averaged eddy viscosity 
and the molecular (kinematic) viscosity is $\langle \nu_{sgs} \rangle/\nu\sim 1$
in the region of interest. For comparison, in \cite{gv:09} this ratio was
$\langle \nu_{sgs} \rangle/\nu\sim 6$.

\begin{figure}[th]
\begin{center}
\includegraphics*[width=0.5\columnwidth,keepaspectratio]{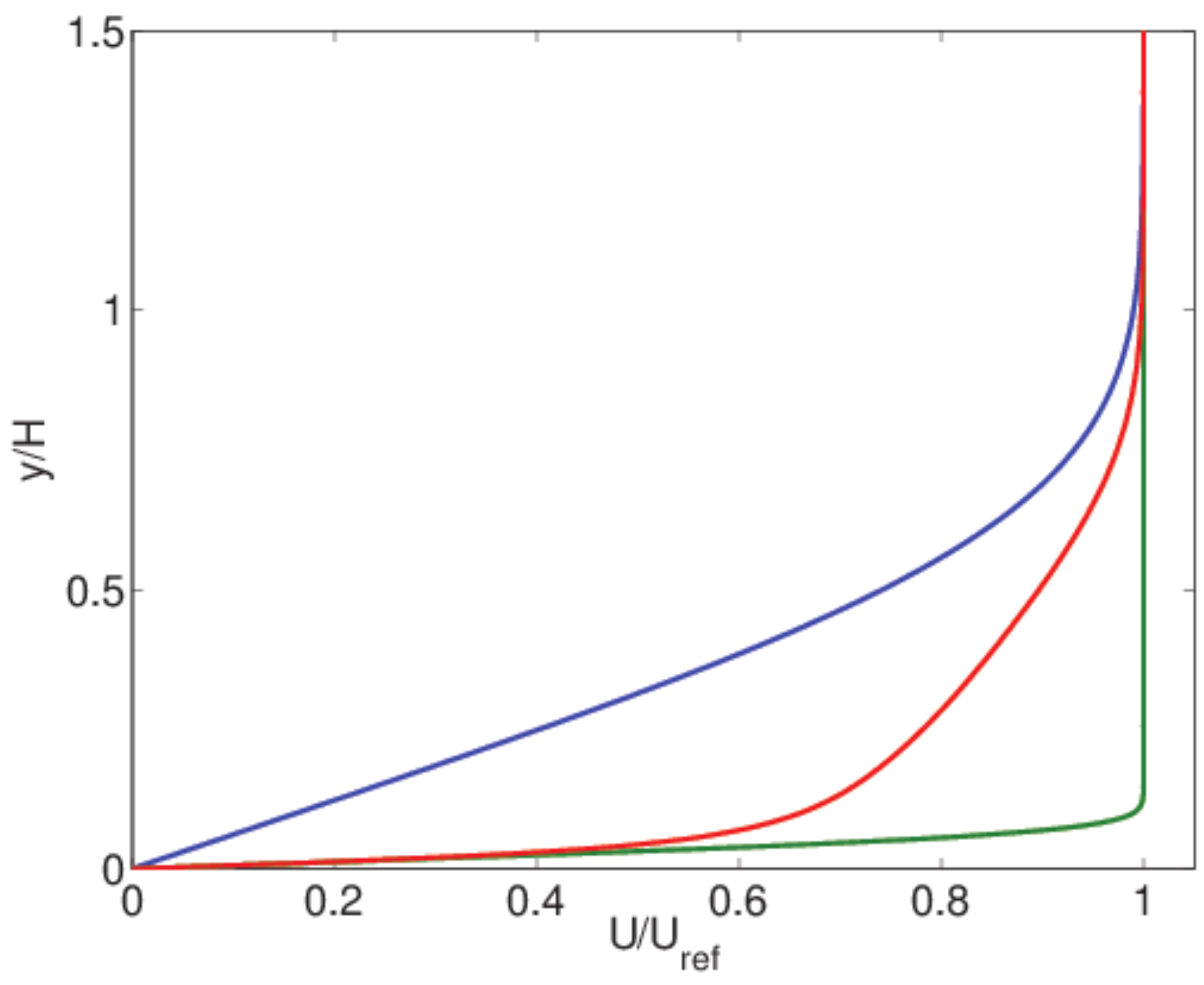}

%
\end{center}
\caption{Profiles of the mean streamwise velocity at the inlet. Line colors defined in Table \ref{tab:1}.}
\label{fig:inlet}
\end{figure}

 A no-slip condition is employed at the bottom wall 
while a free-slip condition is employed at the top boundary. 
Free-slip conditions are also used at the lateral boundaries and convective conditions 
are employed at the exit boundary.
In two of the simulations the inflow conditions are  
time-independent and no turbulence is added to them. 
In these two cases, a Blasius profile is imposed
at the inflow plane. In Simulation S1, $\delta/H=1$ and in Simulation S2,
$\delta/H=0.1$.
In Simulation S3, the approaching boundary layer is turbulent. The mean 
inflow profile corresponds to one of the cases reported in \cite{spalart:88}.
As in \cite{gv:09}, the time-dependent inflow conditions
 are obtained by performing simultaneously
a separate periodic LES of channel flow in which the mean velocity is forced to assume
the desired vertical distribution using a body-force technique \cite{pierce:01}. 
By using this technique, the distribution of turbulent stresses obtained
at the inflow plane is very similar to the standard distribution \cite{spalart:88}
in a fully developed  turbulent boundary layer.
In this precursor calculation, the  number of cells
in streamwise direction is 72. The cost of this simulation is, therefore, $1/8$ of the
total cost. The three inflow profiles are shown in Fig. \ref{fig:inlet} (in the case of Simulation S3 
the mean inflow profile is shown). The Reynolds number of the incoming boundary layer 
based on the momentum 
thickness, $Re_\theta$, for the three cases is also provided in Table \ref{tab:1}.

\begin{table}
\begin{center}
\begin{tabular}{cccccl}
\hline
Case & Inflow & $\delta/H$  & $Re_\theta$  & $C_D$ & Line color\\ \hline
S1  & Laminar & 1           & 900    & 0.305 & Blue \\
S2   & Laminar & 0.1        & 90     & 0.493 & Green\\
S3   & Turbulent & 1        & 670    & 0.407 & Red\\\hline 
\end{tabular}
\end{center}
\caption{Parameters of the simulations. }
\label{tab:1}
\end{table}

%

\section{Results}

After discarding initial transients, statistics have been collected  for a time span of roughly 250 $H/U_{ref}$. This 
corresponds approximately to 11 flow-through times of the computational domain.

\subsection{Pressure distribution}

The mean drag coefficient of the hill, $C_D=D/(0.5\rho U_{ref}^2 S)$, where $D$ is the drag
including pressure and viscous terms, $\rho$ is the fluid density, and $S$ is the frontal surface, 
is reported in Table \ref{tab:1} for the three cases. 
Note that 
using the free stream velocity $U$ in the definition of $C_D$ 
might not be ideal in the present case because of the different
amounts of momentum present in the incoming flow for $y/H<1$ 
(Fig. \ref{fig:inlet}).
The most important contribution to the drag 
is due to the pressure force on the
surface. As an example, the pressure force is responsible for 91\% of the drag in case S3. Similar values are obtained in the other two cases.

\begin{figure}[th]
\begin{center}
\includegraphics*[width=0.6\columnwidth,keepaspectratio]{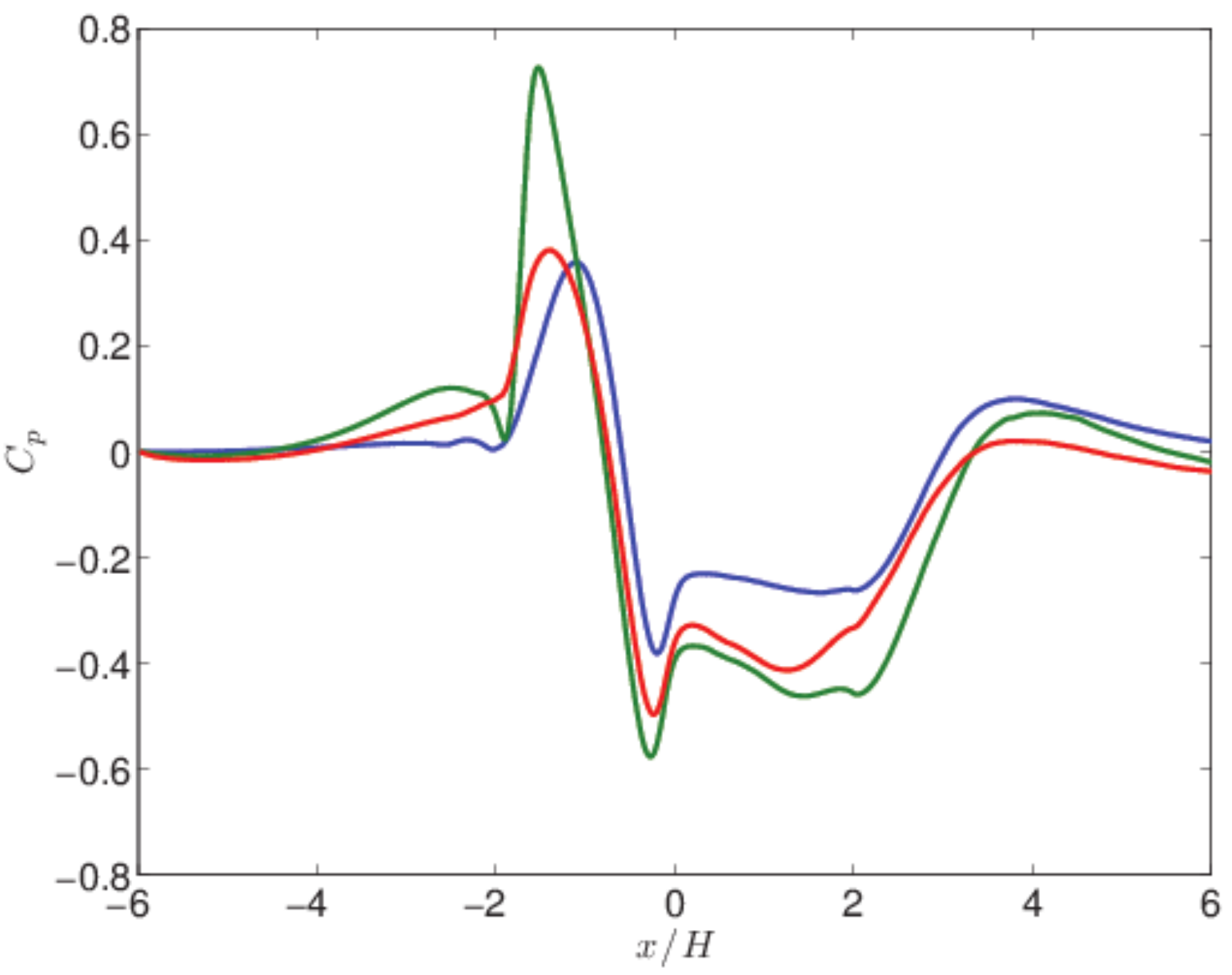}

%
\end{center}
\caption{Pressure coefficient on the bottom surface along the centreline  
as a function of the streamwise coordinate. 
Line colors defined in Table \ref{tab:1}.}
\label{fig:cp}
\end{figure}

Profiles of the pressure coefficient  $C_p=(p-p_\infty)/(0.5 \rho U_{ref}^2)$ 
along the hill centreline
are shown in Fig. \ref{fig:cp}. A similar trend is observed in the three cases. Upstream of the hill the pressure coefficient increases as the hill is approached, reaching a local maximum shortly after
the windward slope of the hill starts. The maximum is more pronounced in case S2, which is the case in which more momentum is present below $y/H=1$ (Fig. \ref{fig:inlet}). As a consequence, this is the case with
the highest drag coefficient of the three.
Thereafter, the flow accelerates and the pressure drops significantly  reaching a local minimum
near the top of the hill. Further downstream,
 the flow decelerates and the pressure recovers somewhat 
but due to the adverse pressure gradient
the flow separates soon, producing a typical plateau in the profile of $C_p$ upto $x/H\sim 2.5$.
The peak pressure at reattachment occurs around $x/H\sim4$ in all cases.

\subsection{Mean flow topology}

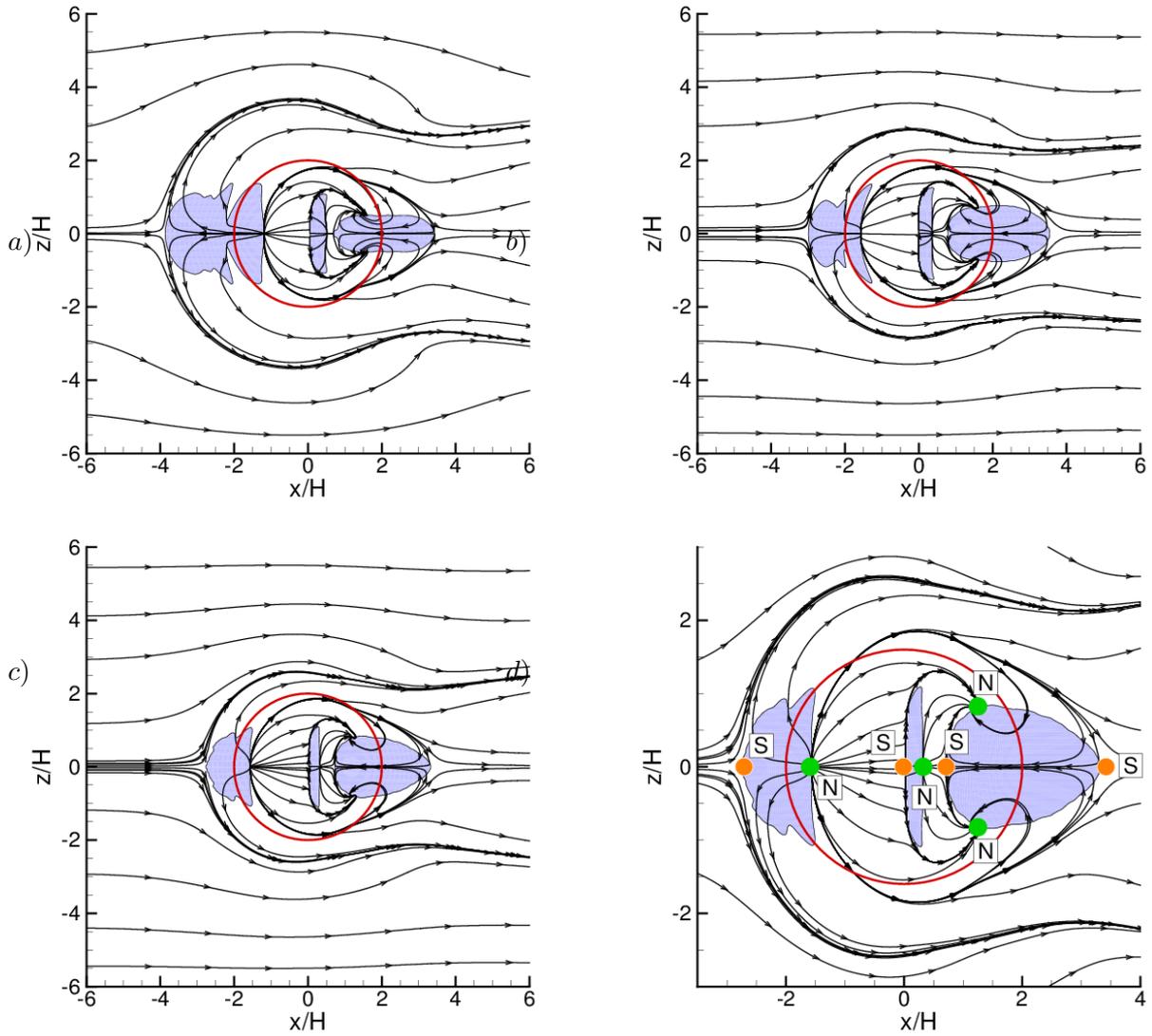
\begin{figure}
\begin{center}
\includegraphics*[width=0.49\columnwidth,keepaspectratio]{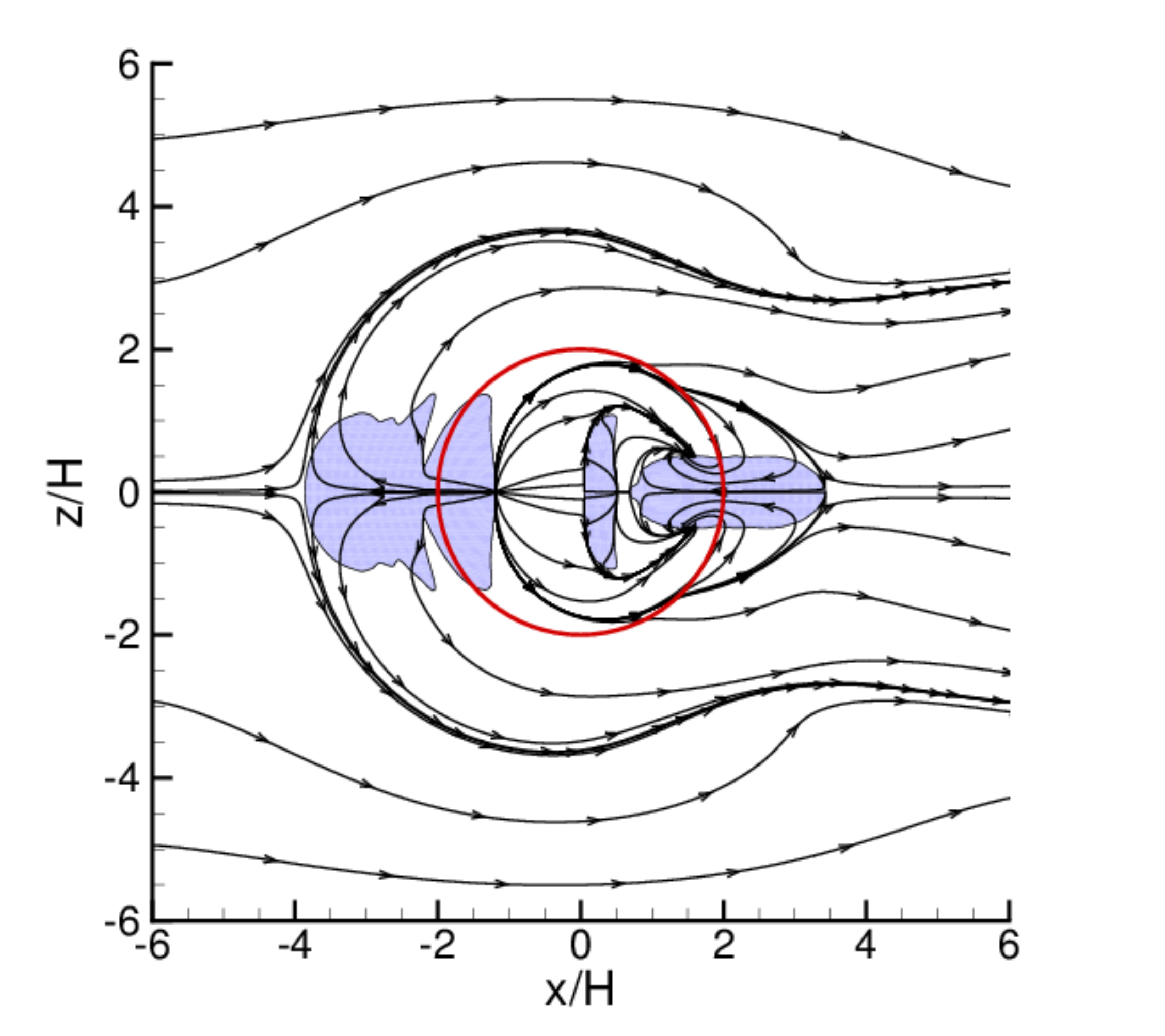}
\includegraphics*[width=0.49\columnwidth,keepaspectratio]{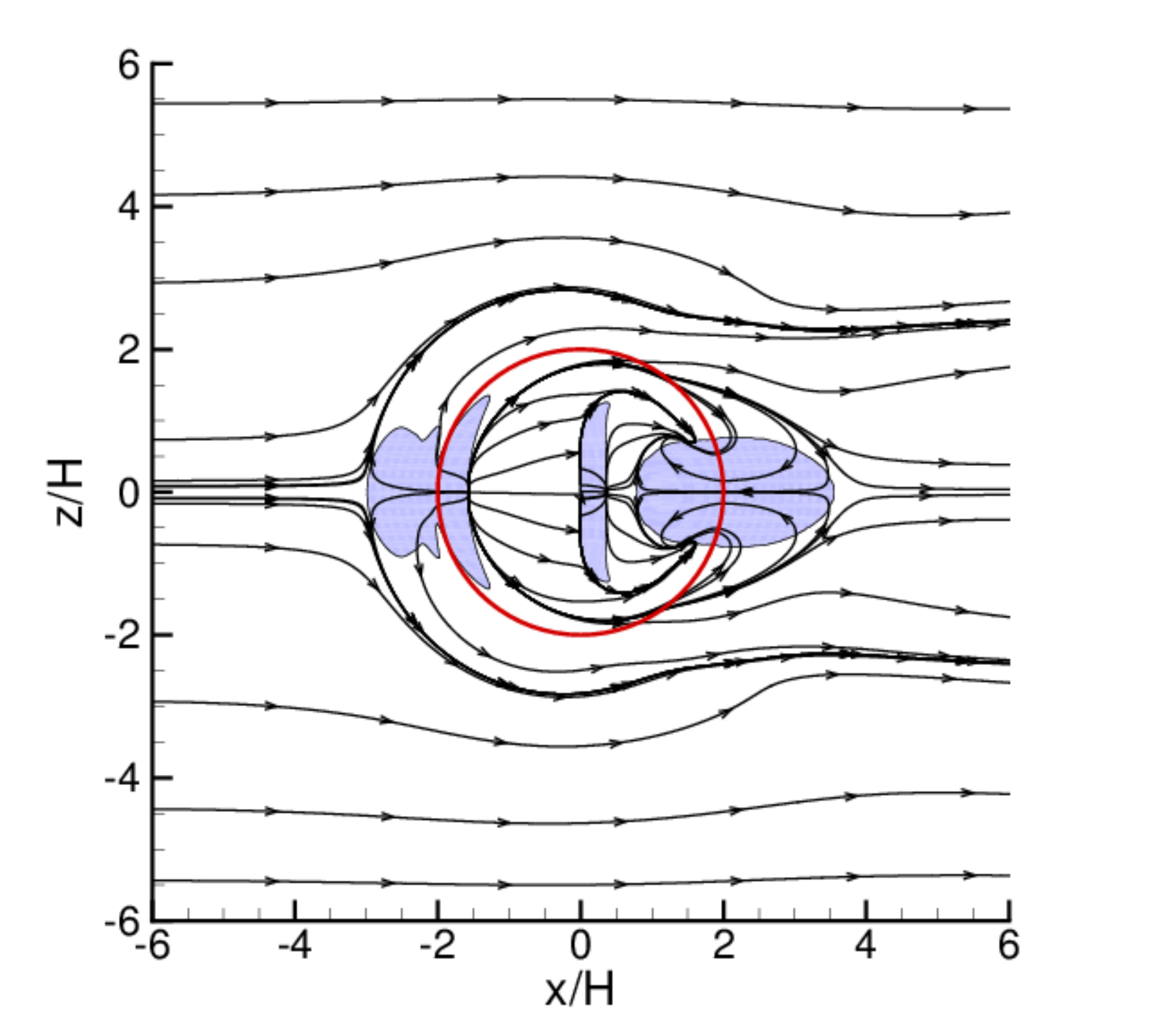}
\includegraphics*[width=0.49\columnwidth,keepaspectratio]{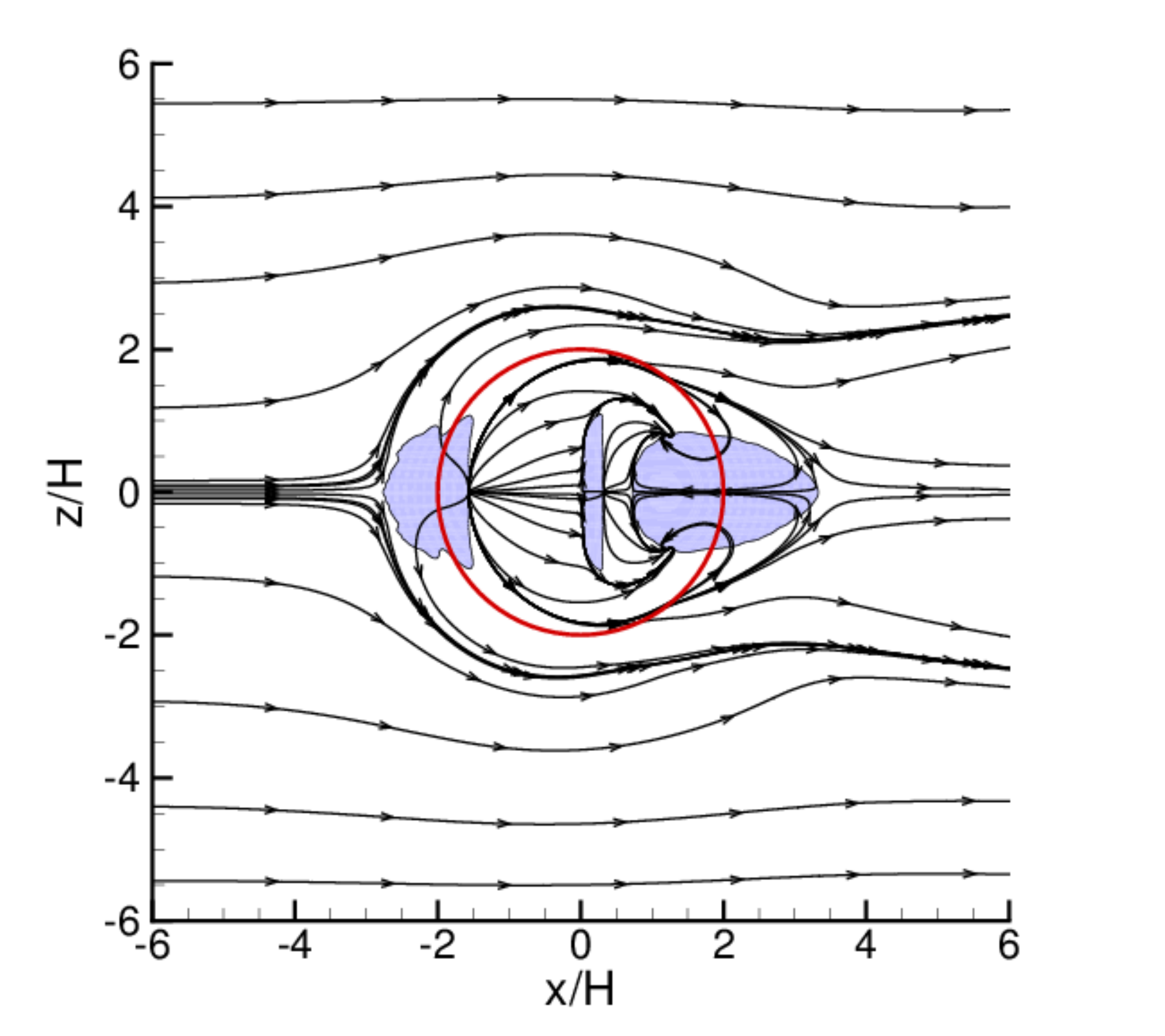}
\includegraphics*[width=0.49\columnwidth,keepaspectratio]{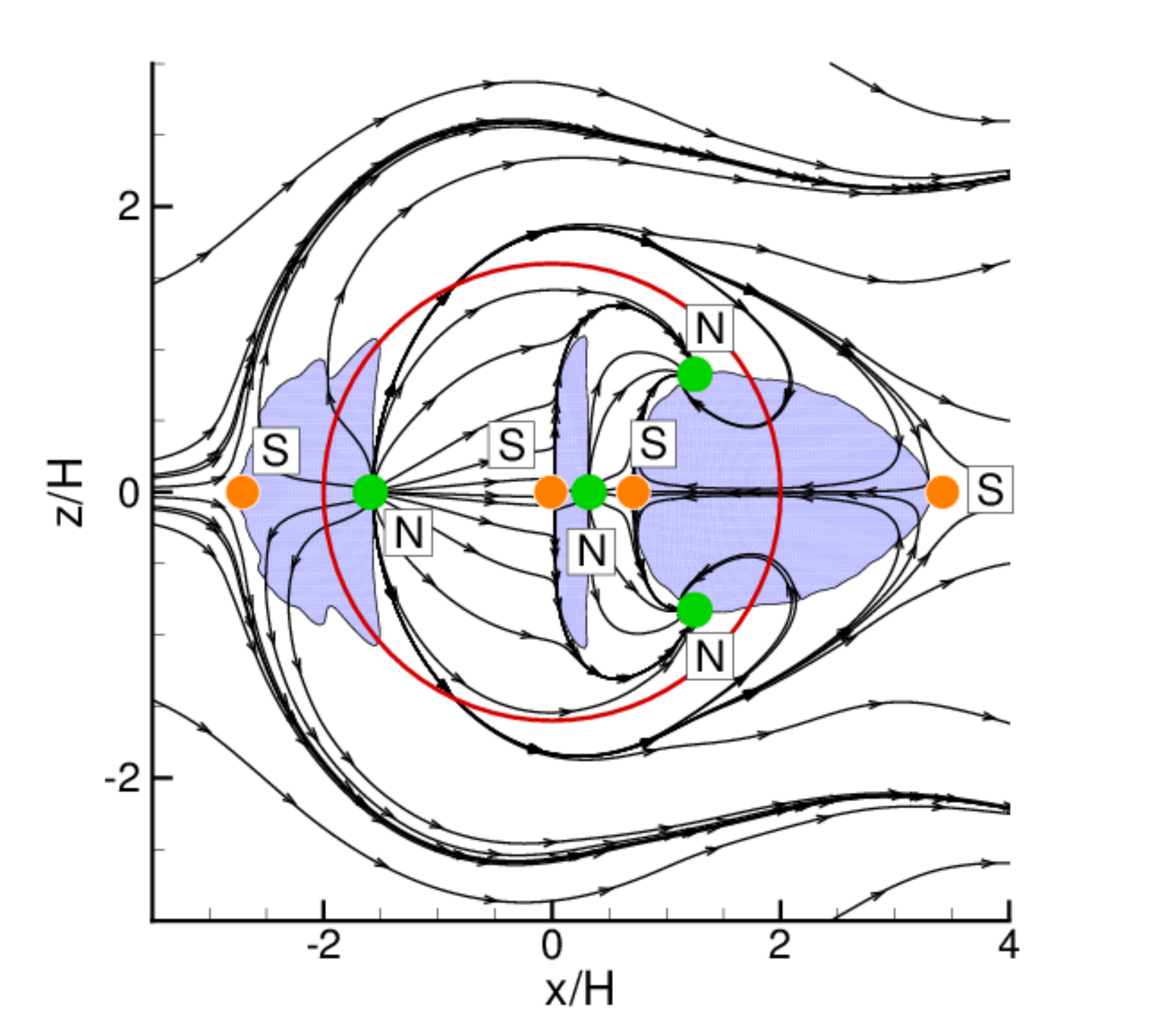}

\end{center}

\begin{picture}(170,0)
\put(1,60){$c)$}
\put(70,60){$d)$}
\put(1,120){$a)$}
\put(70,120){$b)$}
\end{picture}

\caption{Streamlines of the mean flow projected onto a wall-parallel surface 
at a distance to the wall $y/H=0.01$. The blue 
patches indicate the region where backflow is present $U<0$.
a) S1. b) S2. c) S3. d) Zoom of S3 identifying saddles (orange) and nodal points (green).}
\label{fig:str}
\end{figure}

Closely connected with the pressure distribution is the mean flow topology map displayed in 
Fig \ref{fig:str}. This figure shows 
streamlines of the mean flow projected onto a wall-parallel surface 
at a distance to the wall of $y/H=0.01$. In addition, the blue 
patches indicate the region where backflow is present ($U<0$). 
The overall pattern is similar in all three cases but a few differences
are observed.

 A total of eight  topological features can be observed in all
cases (see Fig. \ref{fig:str}$d$ for case S3). 
Four saddle points, which are all
located along the centreline, and four nodes, two located on the centreline, 
and two foci, located on both sides of the hill, around $x/H\sim 1.5$, $z/H\sim \pm 0.5$. As expected, these topological features satisfy the conditions
provided by Hunt et al. \cite{hunt:78} for flow over obstacles (same number of saddle and nodal points). 

In all cases, there are two main areas of backflow, one
in the windward part of the hill and a second one in the rear part. 
The backflow in the windward part of the hill is located between a
saddle point and a nodal point on the centreline. The appearance of this region 
is related to a well-know phenomenon
in the flow around wall-mounted obstacles: 
the formation of a horseshoe vortex at the foot of the
obstacle. This has been observed in flow over wall-mounted cubes, cylinders, etc.  However, this is not observed for the present geometry at
significantly higher Reynolds numbers \cite{gv:09}.
It is noticeable that this region is largest in Simulation S1,
 while in S2 and S3 the smaller regions are of comparable size. The streamlines
arriving at the first saddle point from upstream, are deviated to both sides 
to go around the hill. In the Simulation S1 they are deviated as far as $|z/H|\sim4$
while in S2 and S3 they remain within $|z/H|\sim3$.

The backflow region in the rear part has a more complex structure. It is split 
in two parts with a forward flow region in between. It turns out that this forward 
flow region is extremely thin (as discussed below), 
this is why we refer to the rear backflow region as a single 
region. The differences between the three simulations are minor. The location
of saddle and nodal points are approximately the same in all cases and therefore 
the topology of the streamlines that connect them is the same.

\begin{figure}
\begin{center}
\includegraphics*[width=0.9\columnwidth,keepaspectratio]{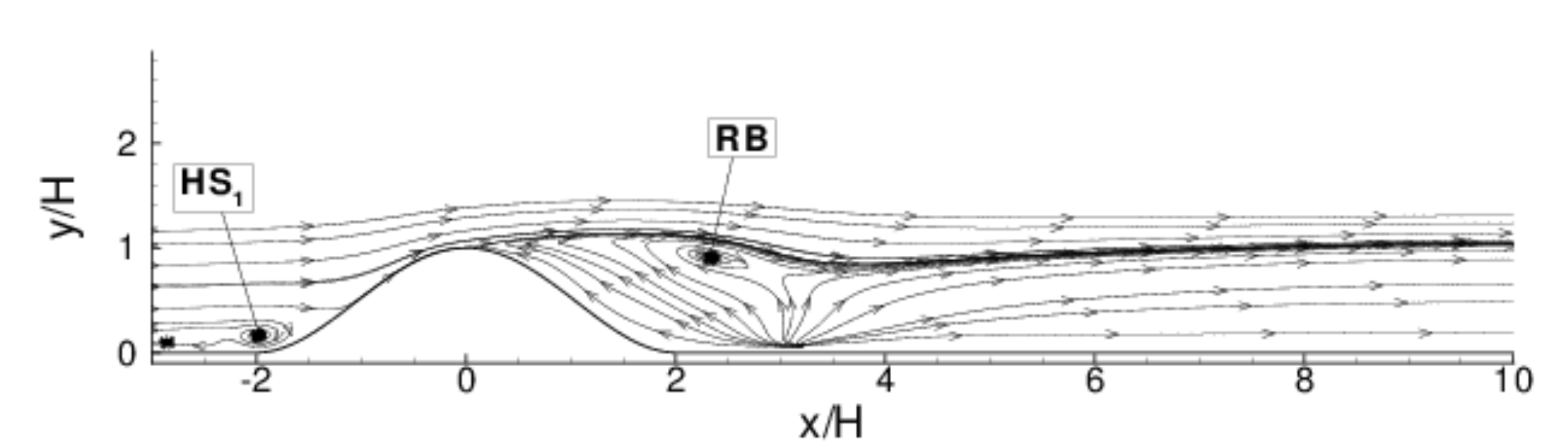}
\includegraphics*[width=0.9\columnwidth,keepaspectratio]{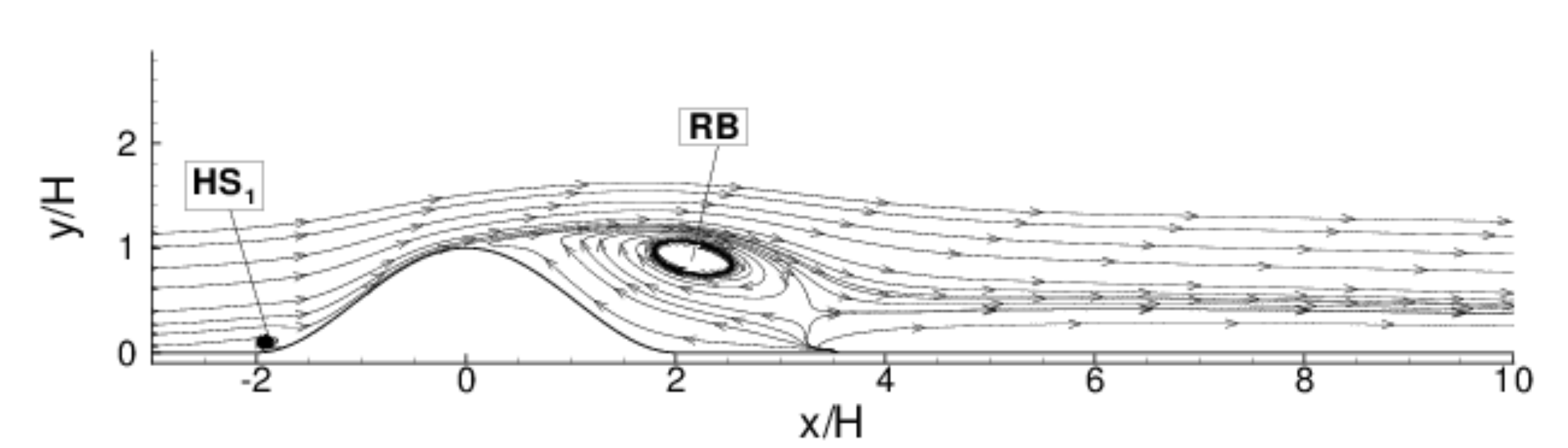}
\includegraphics*[width=0.9\columnwidth,keepaspectratio]{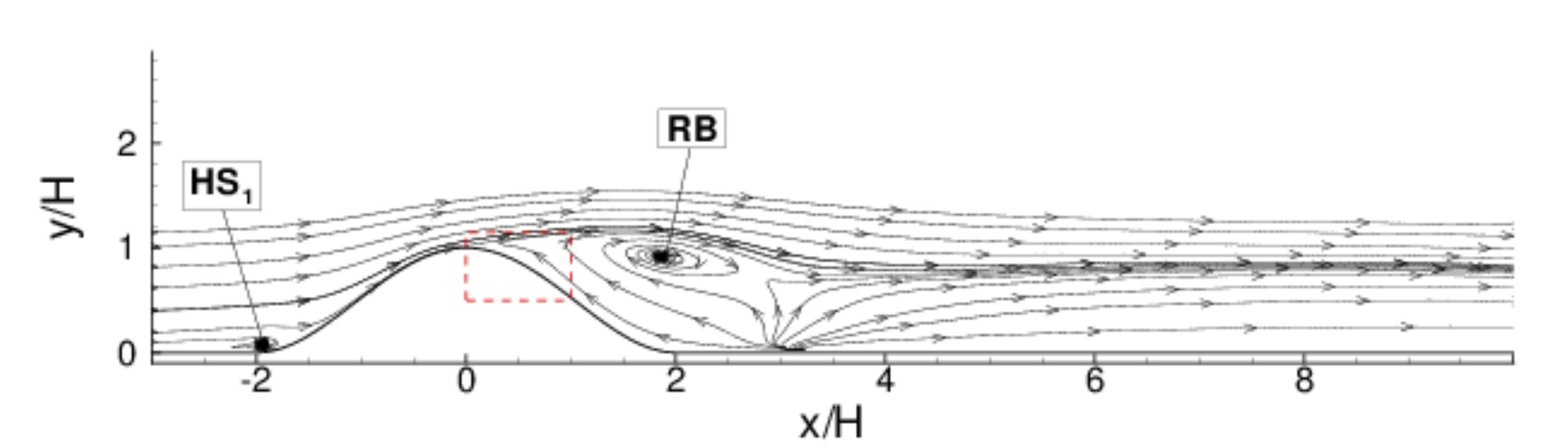}
\end{center}
\caption{Streamlines of the mean flow in the midplane. Top, S1. Middle, S2. Bottom, S3. The red square identifies the zoomed-in region shown in 
Fig. \ref{fig:zoom}}
\label{fig:str_mid}
\end{figure}

\subsection{Mean velocity distribution}

Figure~\ref{fig:str_mid} shows a comparison of streamlines of the mean flow in 
the symmetry plane of Simulations S1, S2 and S3. The main recirculation region behind the hill
is longer and higher than the one observed for this same
geometry at a significantly higher Reynolds number \cite{gv:09}. 
 The centre of the main recirculation 
bubble behind the crest of the hill is identified by the label $RB$. While 
$HS_1$ identifies the small upstream area of recirculation (vortex) obtained at the 
foot of the hill. Because of the strong accelerating mean flow along the foot of the
hill, the upstream vortex $HS_1$ is stretched in the streamwise direction. As it
is wrapped partially around the foot of the hill the horse shoe vortex
mentioned above is formed.
The shape of the horse shoe vortex is reflected in the lateral 
streamlines that originate from the upstream saddle point, shown in
Figure~\ref{fig:str}.
Because of the larger $Re_\theta$ (Tab. \ref{tab:1}) of the incoming (laminar)
boundary layer,
compared to Simulations S2 and S3, in Simulation S1 a significantly 
larger upstream separation bubble $HS_1$ is generated. Also, the streamlines
- originating from the crest of the hill - that bound the wake-like
wall-parallel flow show that the height of the wake increases with increasing
$Re_{\theta}$ of the inflow profile (see Table~\ref{tab:1}). 

\begin{figure}
\begin{center}
\includegraphics*[width=0.5\columnwidth,keepaspectratio]{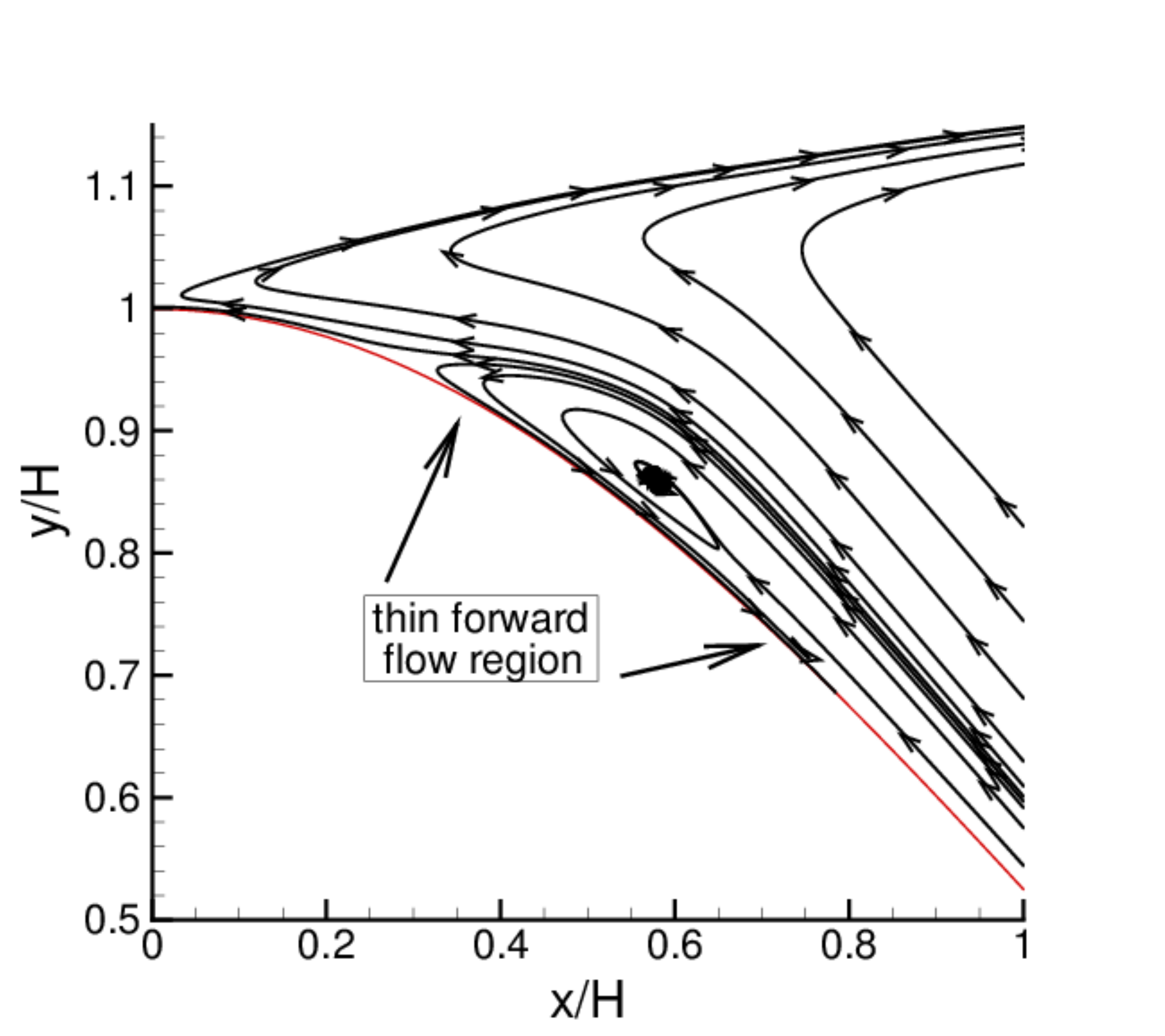}
\end{center}
\caption{Streamlines of the mean flow in the midplane (Zoomed view). Simulation S3.}
\label{fig:zoom}
\end{figure}

Below the main recirculation region, a very shallow secondary bubble is obtained in all three simulations. Evidence of this can be seen in Fig. \ref{fig:str}. However, this bubble is not visible in Fig. \ref{fig:str_mid}.
A zoomed view of simulation S3, presented in Fig. \ref{fig:zoom}, 
illustrates the shape of the secondary bubble. 
This thin region is resolved in the simulation
with 8 to 10 grid points in wall-normal direction.

\begin{figure}
\begin{center}
\includegraphics*[width=\columnwidth,keepaspectratio]{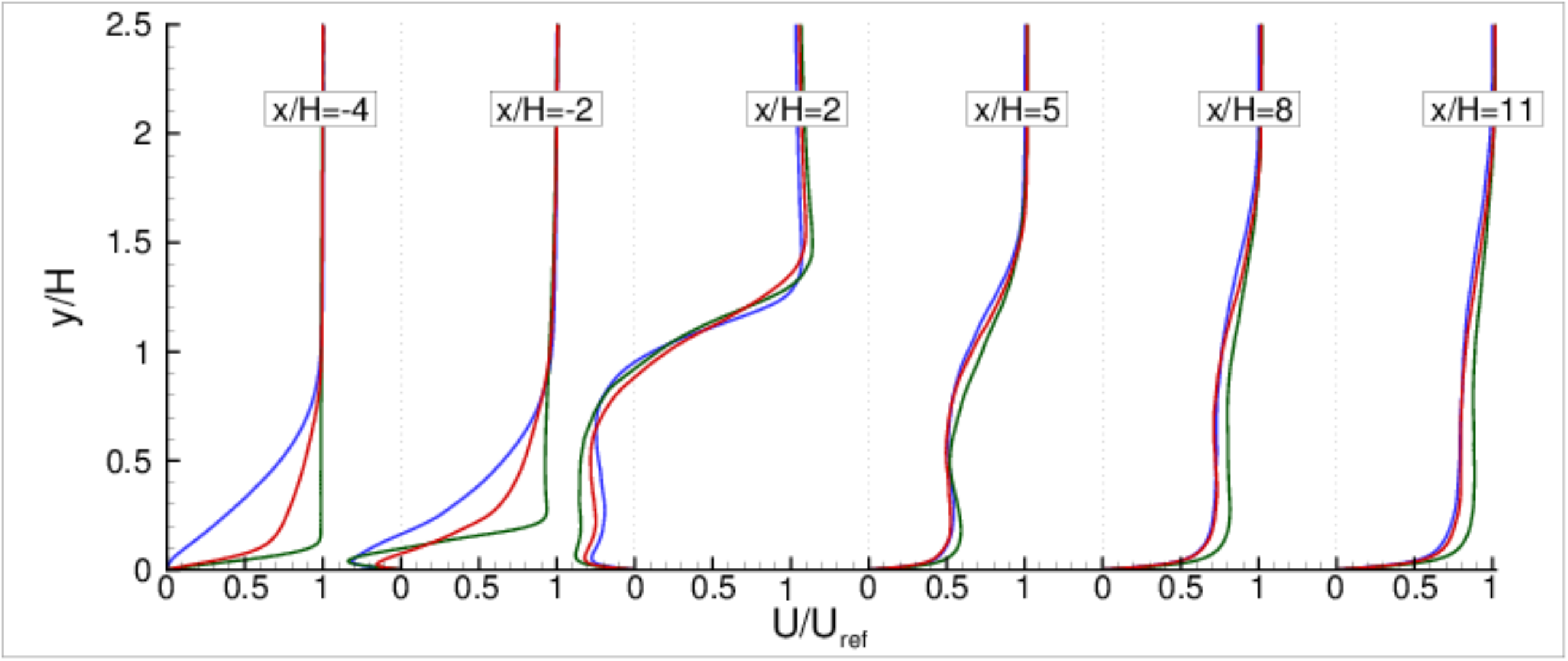}
\end{center}
\caption{Profiles of mean streamwise velocity in the midplane $z/H=0$, 
at various streamwise locations: $x/H=$-4, -2, 2, 5, 8, 11.
Line colors defined in Table \ref{tab:1}.}
\label{fig:ut_comp}
\end{figure}

\par
Figure~\ref{fig:ut_comp} shows the streamwise velocity profiles of the three
simulations at $x/H=-4,\,-2,\,2,\,5,\,8,\,11$. At $x/H=-2$ the profiles confirm
the presence of the separation bubble $HS_1$ at the foot of the hill in all
simulations. The presence of the main
recirculation bubble $RB$ is clearly reflected in all three simulations 
by the reverse flow in the profiles shown at $x/H=2$. Despite the difference 
in the profiles and the state (laminar/turbulent) of the flow at the inflow 
plane, for $x/H = 2, \ldots, 11$ 
the mean profiles of all three simulations do not present significant differences. 
Furthermore, the profiles for $x/H=5,8,11$ all exhibit the characteristic full 
turbulent boundary layer profile at the bottom with a wake-like region on top.

\begin{figure}
\begin{center}
\includegraphics*[width=\columnwidth,keepaspectratio]{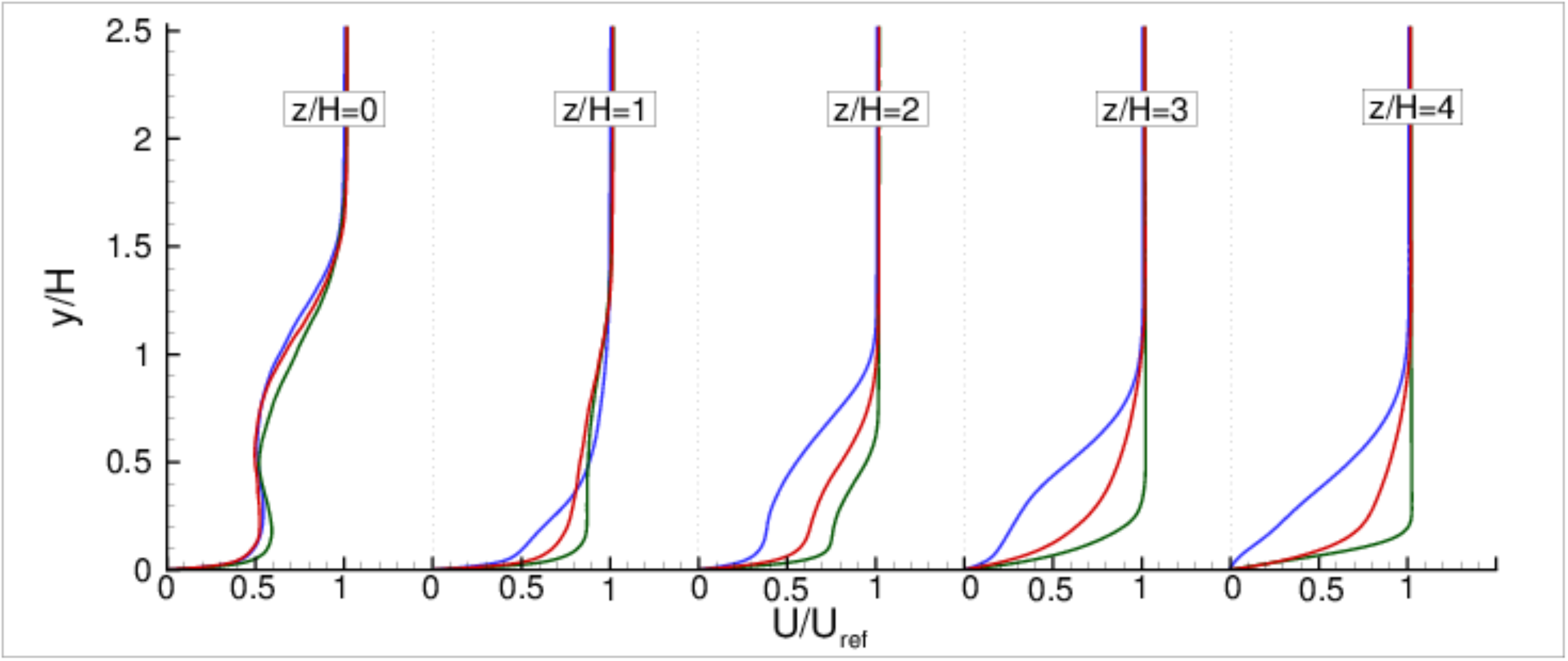}
\end{center}
\caption{Profiles of mean streamwise velocity at $x/H=5$, 
at various spanwise locations: $z/H=$0, 1, 2, 3, 4.
Line colors defined in Table \ref{tab:1}.}
\label{fig:ut_compx5}
\end{figure}

Figure~\ref{fig:ut_compx5} shows a comparison of the mean $u$-velocity profiles 
at various stations $z/H=0,1,2,3$ and $4$, extracted in the cross section 
at $x/H=5$. While at the symmetry plane ($z/H=0$) the velocity profiles almost 
collapse, towards the edges of the computational domain gradually more and more
differences can be observed. At $z/H=4$, finally, for each simulation the 
shape of the mean $u$-velocity profile is found to be very similar 
to the mean inflow velocity profile. Hence,  
the downstream influence of the hill 
(in the form of a wake) is only noticeable 
in a spanwise region of limited size. 

\begin{figure}
\begin{center}
\includegraphics*[width=0.9\columnwidth,keepaspectratio]{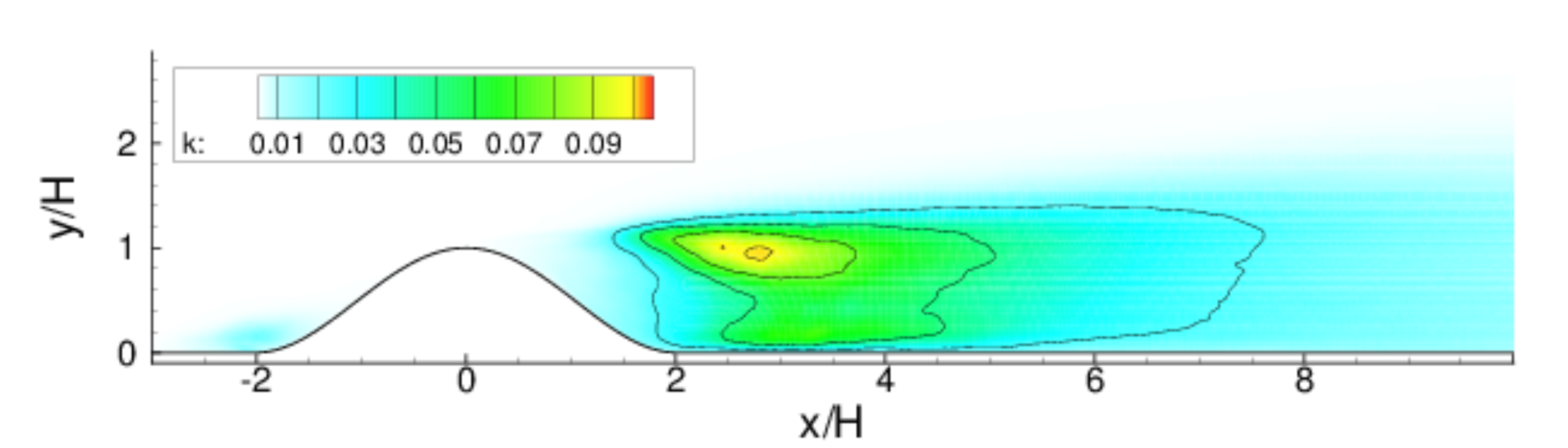}
\includegraphics*[width=0.9\columnwidth,keepaspectratio]{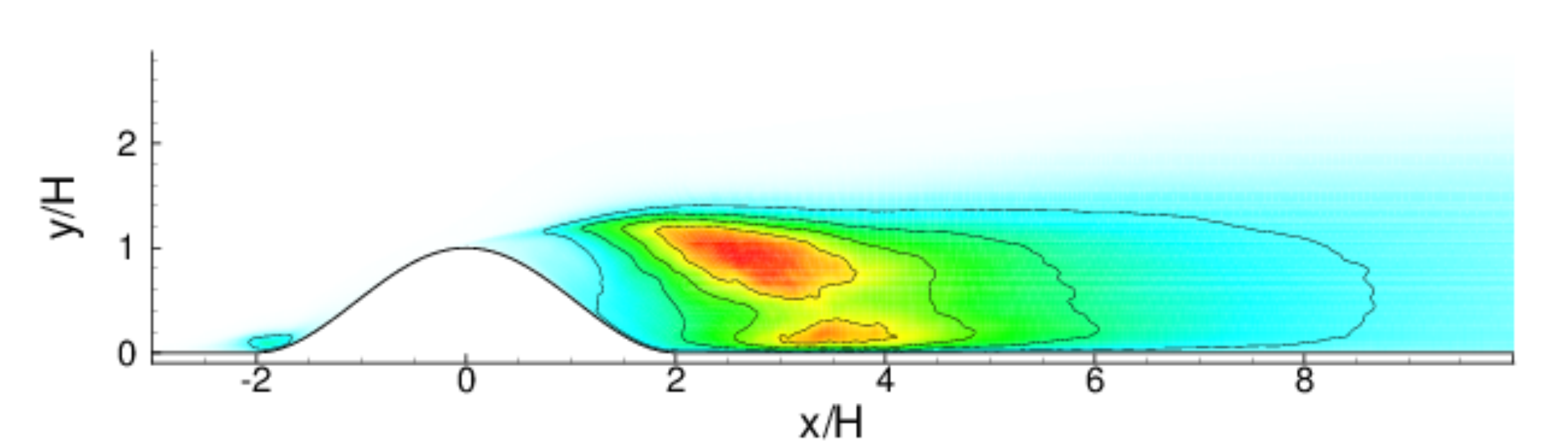}
\includegraphics*[width=0.9\columnwidth,keepaspectratio]{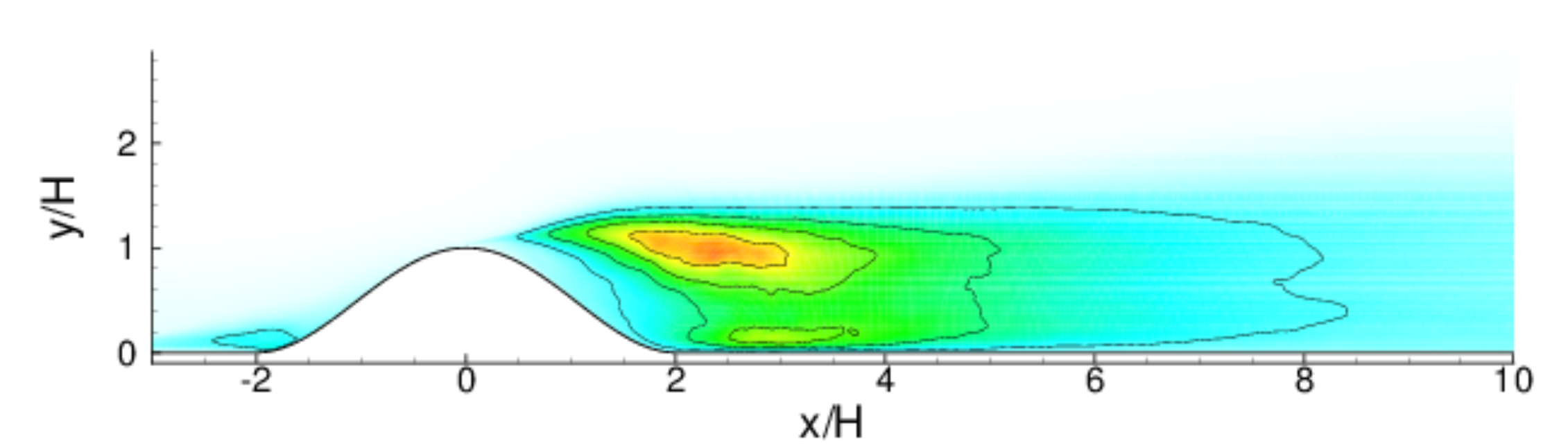}

\end{center}
\caption{Contours of turbulent kinetic energy in the midplane. Top, S1. Middle, S2. Bottom, S3. Lines are $k/U_{ref}^2=$0.025,0.05, 0.075 and 0.1.}
\label{fig:k_contours}
\end{figure}

\subsection{Turbulence in the symmetry plane}

Figure~\ref{fig:k_contours} shows the turbulent kinetic energy, $k$, 
in the symmetry plane
 for Simulations S1, S2 and S3. Immediately upstream of the
hill, at $x/H \approx -2$, in each simulation a small patch with an elevated $k$-level can be
observed that coincides with the upstream separation bubble labelled $HS_1$
in Figure~\ref{fig:str_mid}. The production of kinetic energy
leading to increased values of $k$
 in the re-circulation
zone of a separation bubble was observed earlier in~\cite{wissink:03,
wissink:06} and was accounted for by an elliptic instability of the rolled-up 
shear layer.
Downstream of the small separation bubble, the streamwise pressure 
gradient turns favourable and the energized boundary layers in all simulations 
re-attach.  
Also, at the centre of the circulation bubble, downstream of the 
crest of the hill - labelled $RB$ in Figure~\ref{fig:str_mid} - production of kinetic
energy is observed in all simulations. 
Of the three simulations, the momentum thickness of the incoming boundary layer 
in Simulation $S2$ is smallest, followed by Simulation $S3$ and then 
Simulation $S1$. Because of this, in the inflow region the wall-shear in Simulation B is
significantly larger than in Simulations S1 and S2.  
As the boundary layer separates from the crest
of the hill, the mean shear in the free-shear layers generates 
turbulence. Because in Simulation S2 the mean shear is much stronger than 
in Simulations S1 and S3, the production of $k$ in Simulation S2 is significantly 
higher (as confirmed in Figure~\ref{fig:k_contours}). 
Similarly, because of the difference
in wall-shear strength, in Simulation S1 the production of $k$ is found to be
lower than in Simulation S3. 
\par
\begin{figure}
\begin{center}
\includegraphics*[width=\columnwidth,keepaspectratio]{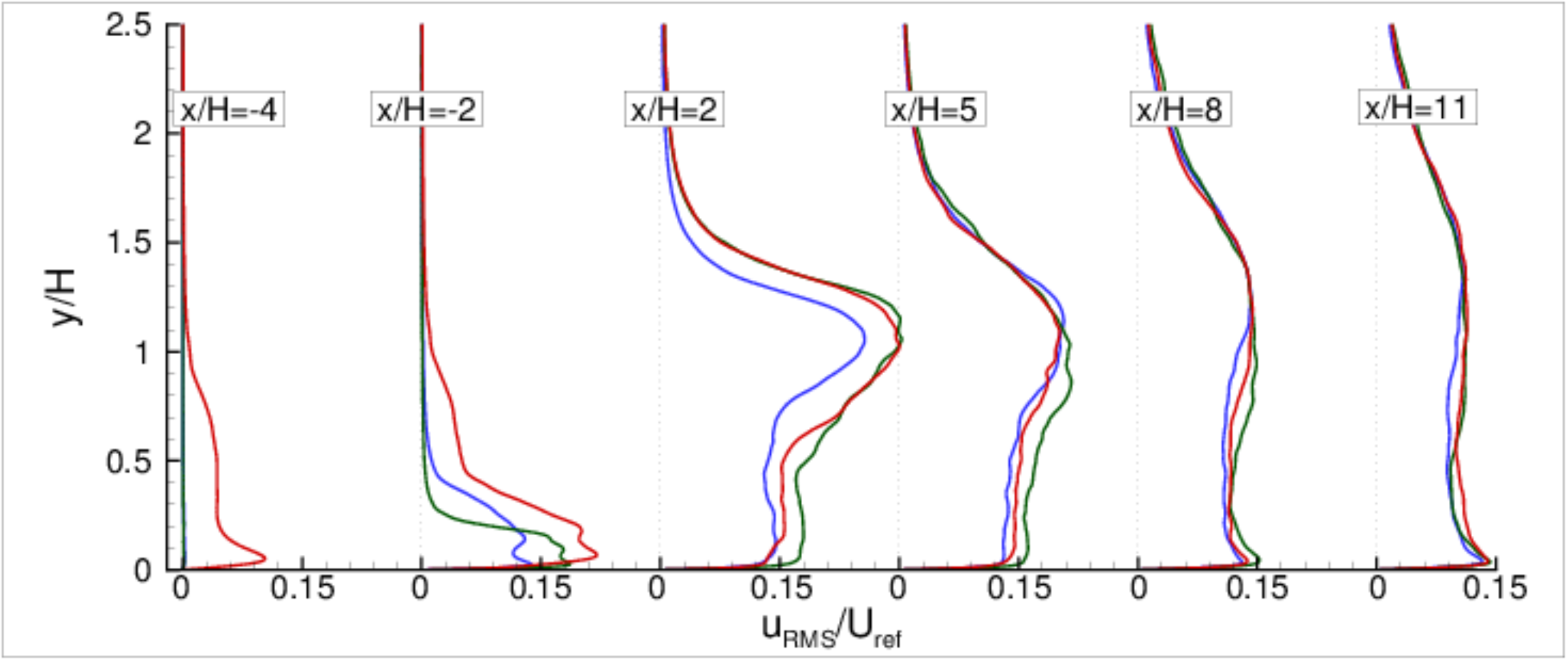}
\includegraphics*[width=\columnwidth,keepaspectratio]{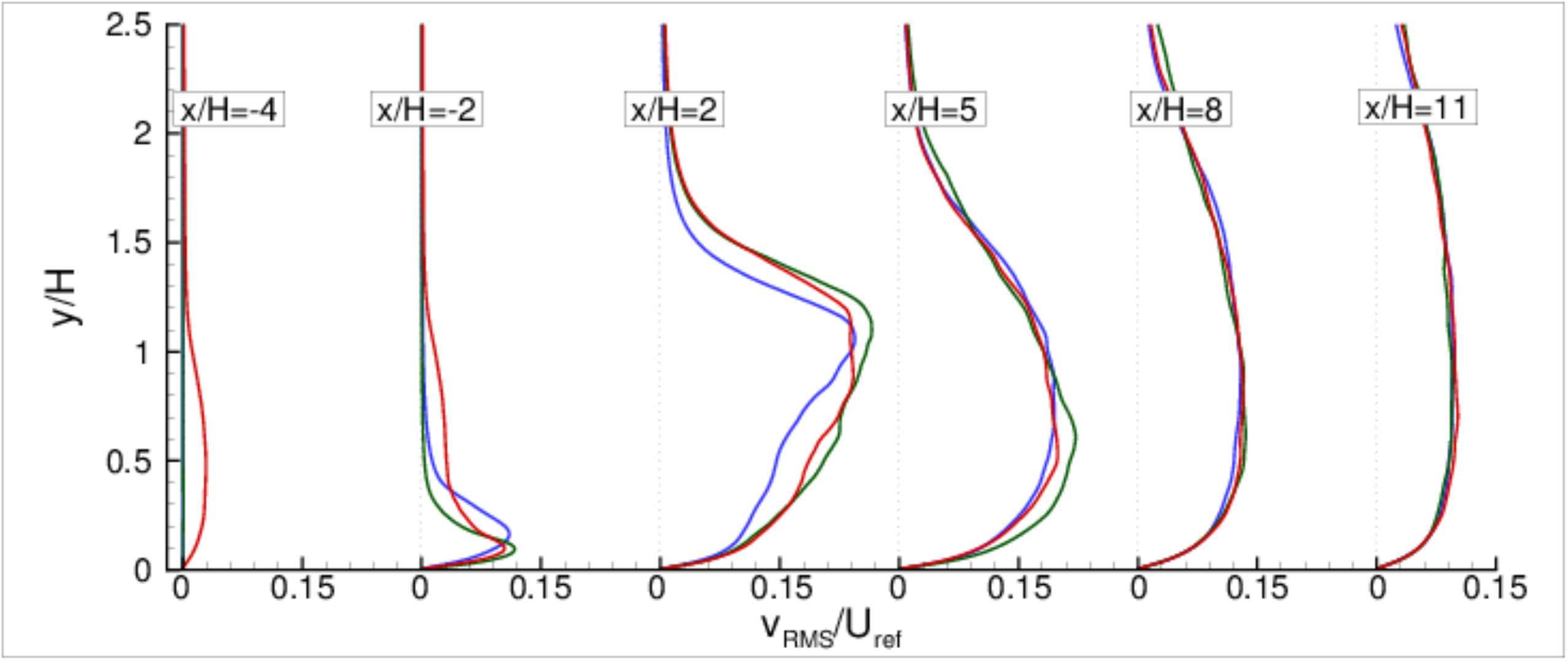}
\includegraphics*[width=\columnwidth,keepaspectratio]{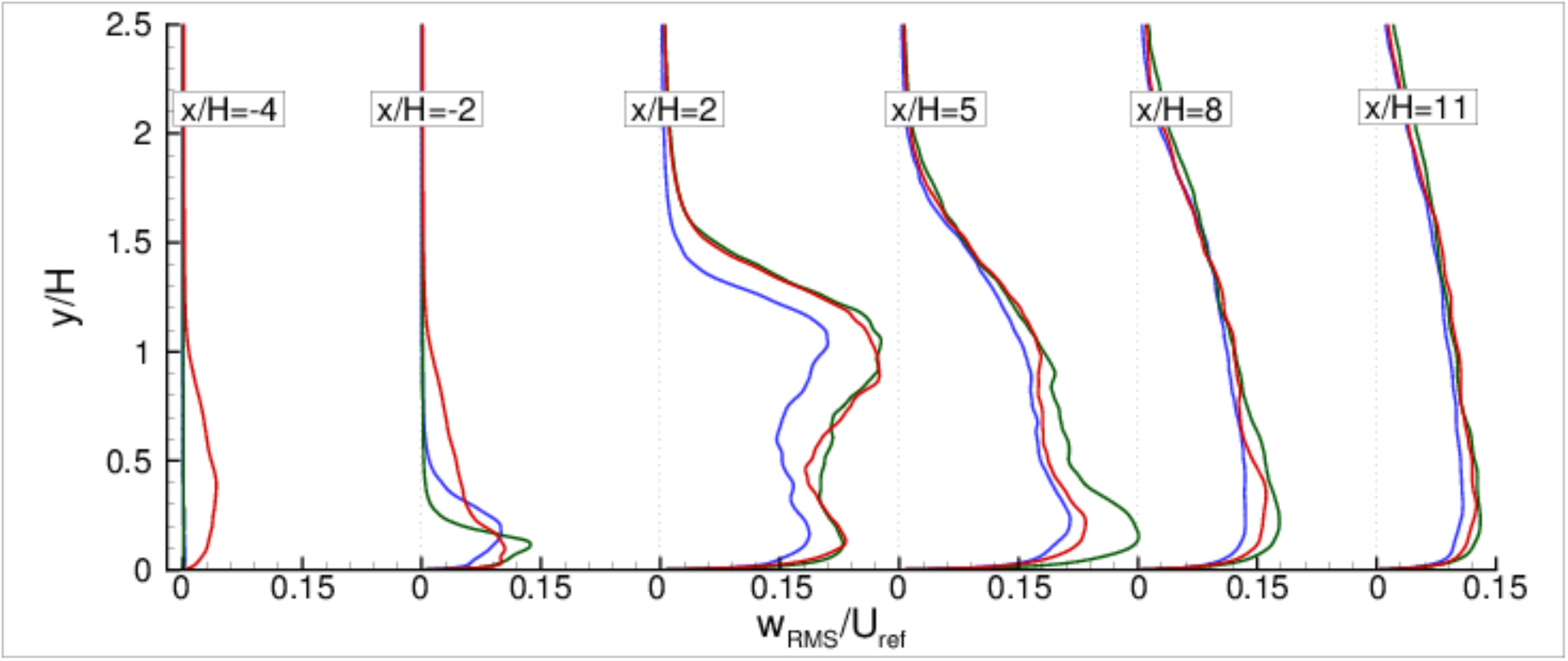}
\end{center}
\caption{Profiles of rms-velocity components in the midplane $z/H=0$, 
at various streamwise locations: $x/H=$-4, -2, 2, 5, 8, 11.
Top, streamwise velocity $u_{rms}/U_{ref}$. 
Middle, vertical velocity $v_{rms}/U_{ref}$. 
Bottom, spanwise velocity $w_{rms}/U_{ref}$.
Line colors defined in Table \ref{tab:1}.}
\label{fig:rms_comp}
\end{figure}

Profiles of the rms values of the $u$-, $v$- and $w$-velocity components of the
three simulations are shown in Figure~\ref{fig:rms_comp}. Close to the inflow plane, 
at $x/H=-4$, the rms-values of all velocity components of the two 
Simulations S1 and S2 are observed to be zero. In the upstream separation bubble,
$HS_1$, at $x/H=-2$, all three velocity components exhibit fluctuations as
reflected by the locally increased rms-values. On the lee side of the hill, at
$x/H=2$, the presence of the large recirculation bubble $RB$ induces significant
fluctuations in all simulations. 
Because of the increased momentum thickness of Simulation
S1's incoming boundary layer,the fluctuation in the
rms-values is found to be slightly less than in 
Simulations S2 and S3. For $x/H \ge 5$, the profiles of the 
rms values of the velocity components become similar, indicating that 
the significant differences in the shape and state (laminar/turbulent) of the 
boundary layers at the inflow plane is no longer identifyable by studing
differences in the rms values in the symmetry plane. 
Compared to the Simulations S1 and S3, the 
only significant difference is observed in the $w_{rms}$ values of Simulation S2 
at $x/H=5$ which, close the the lower wall, are significantly higher than in
Simulations S1 and S3. This is likely related to the increased mass flux  
in Simulation S2.  

\subsection{Secondary motions}

\begin{figure}
\begin{center}
\includegraphics*[width=0.32\columnwidth,keepaspectratio]{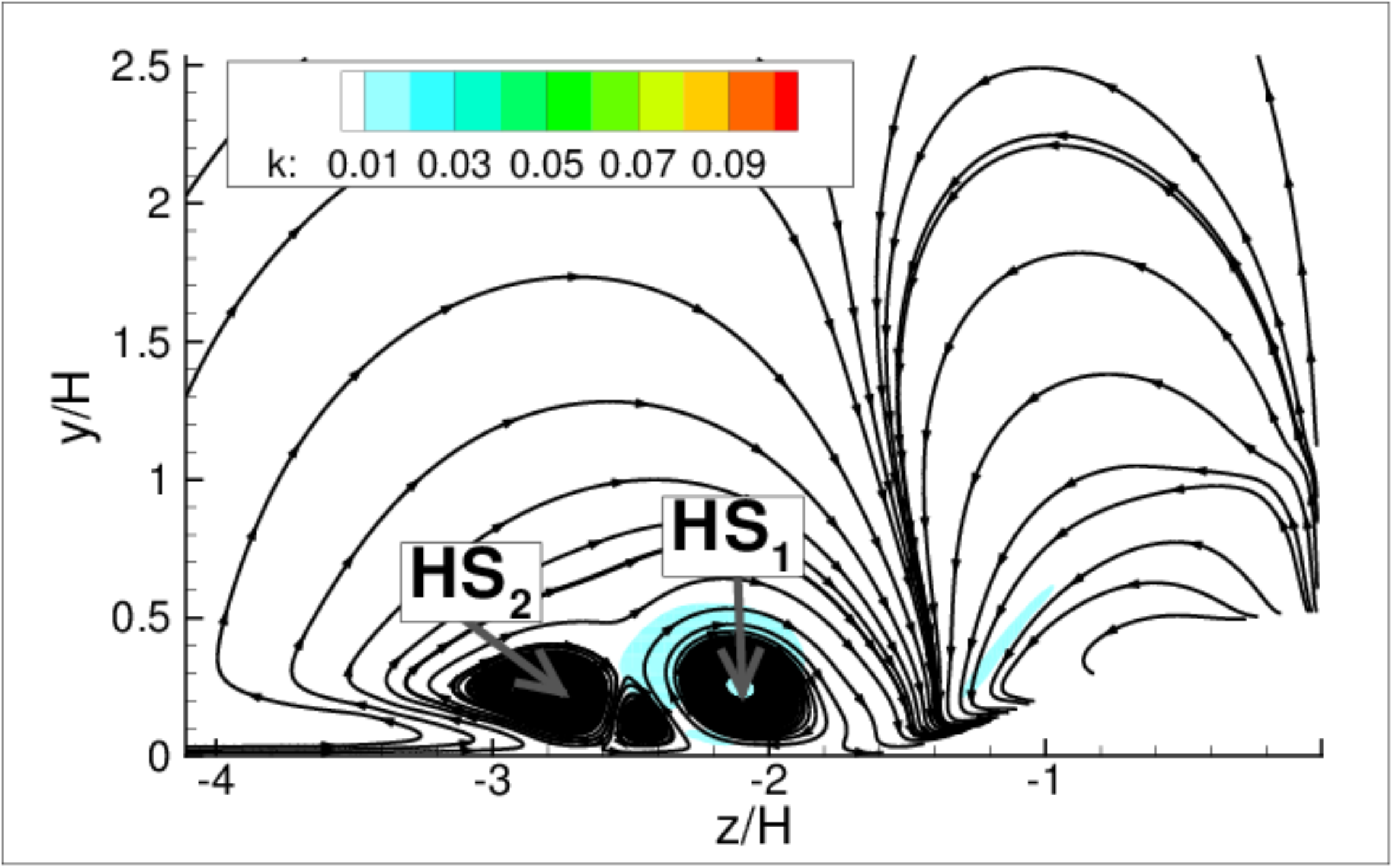}
\includegraphics*[width=0.32\columnwidth,keepaspectratio]{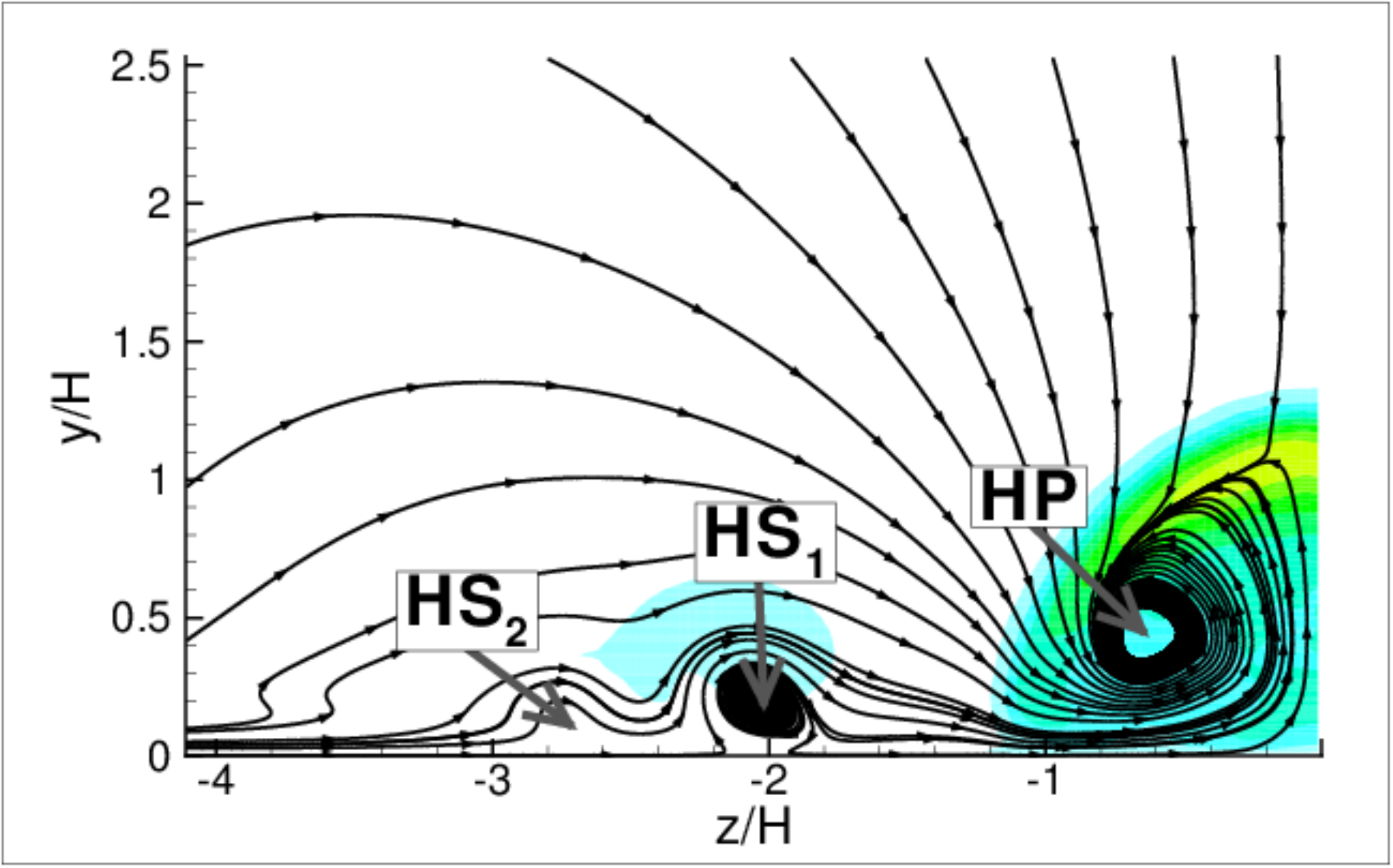}
\includegraphics*[width=0.32\columnwidth,keepaspectratio]{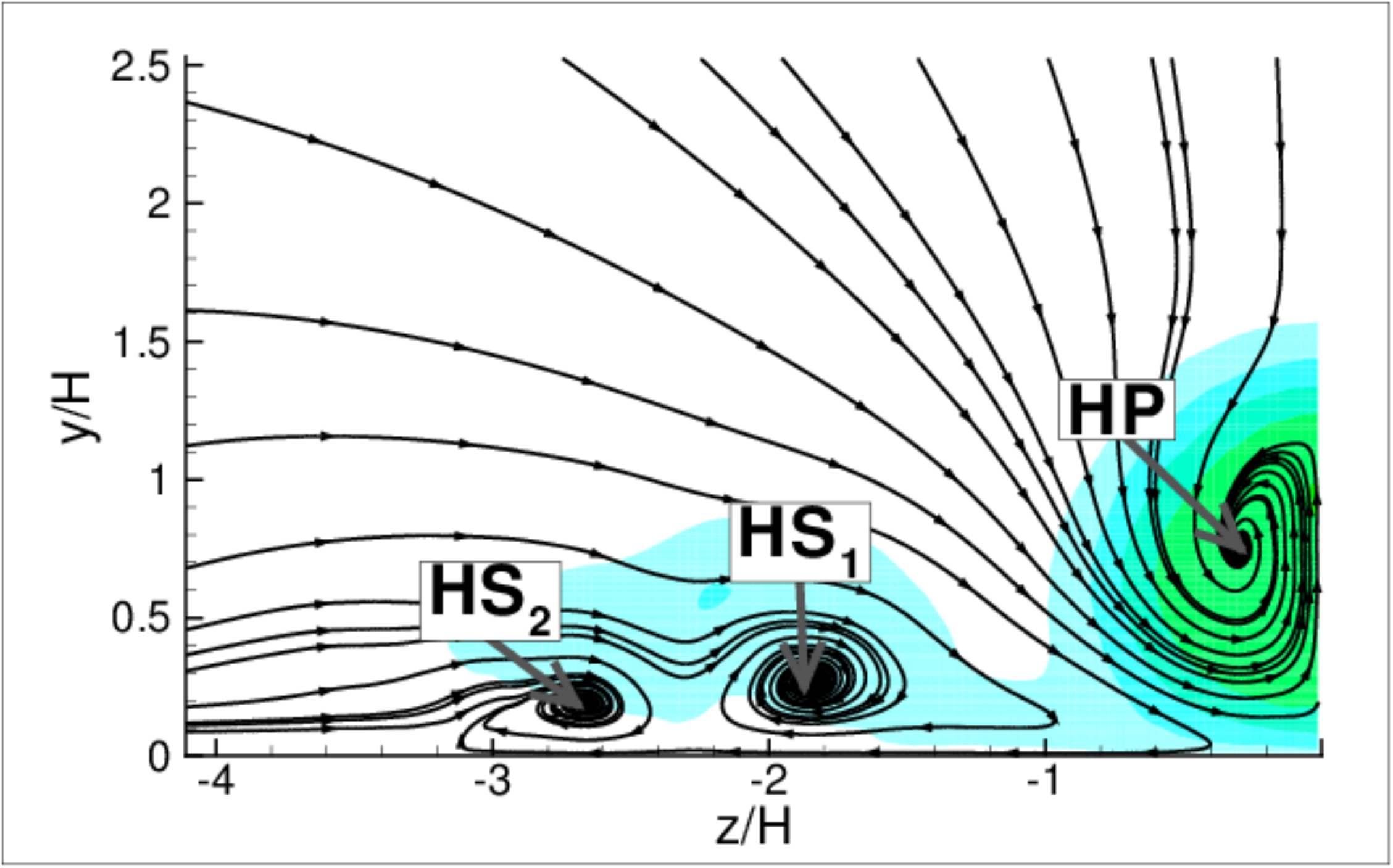}
\includegraphics*[width=0.32\columnwidth,keepaspectratio]{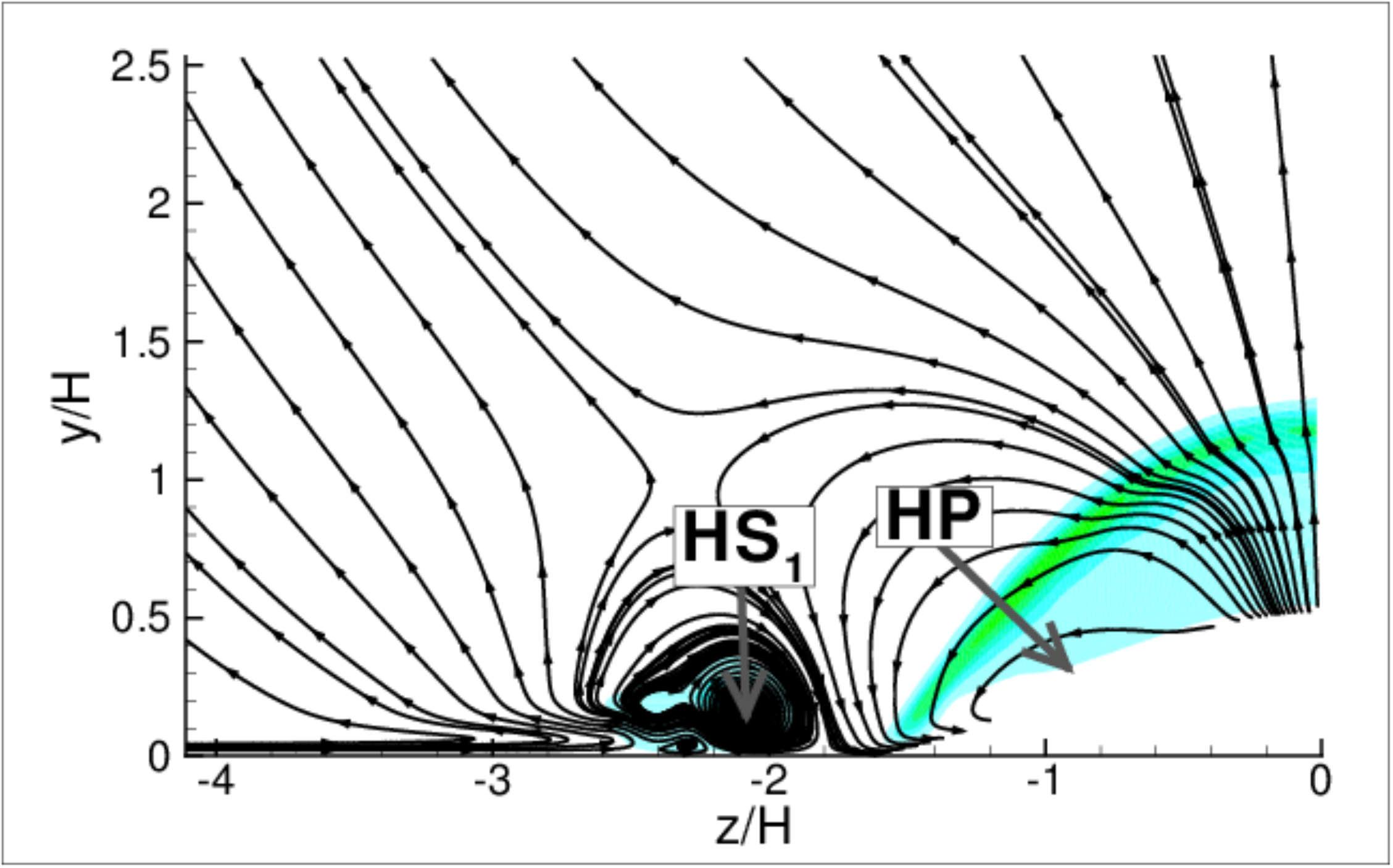}
\includegraphics*[width=0.32\columnwidth,keepaspectratio]{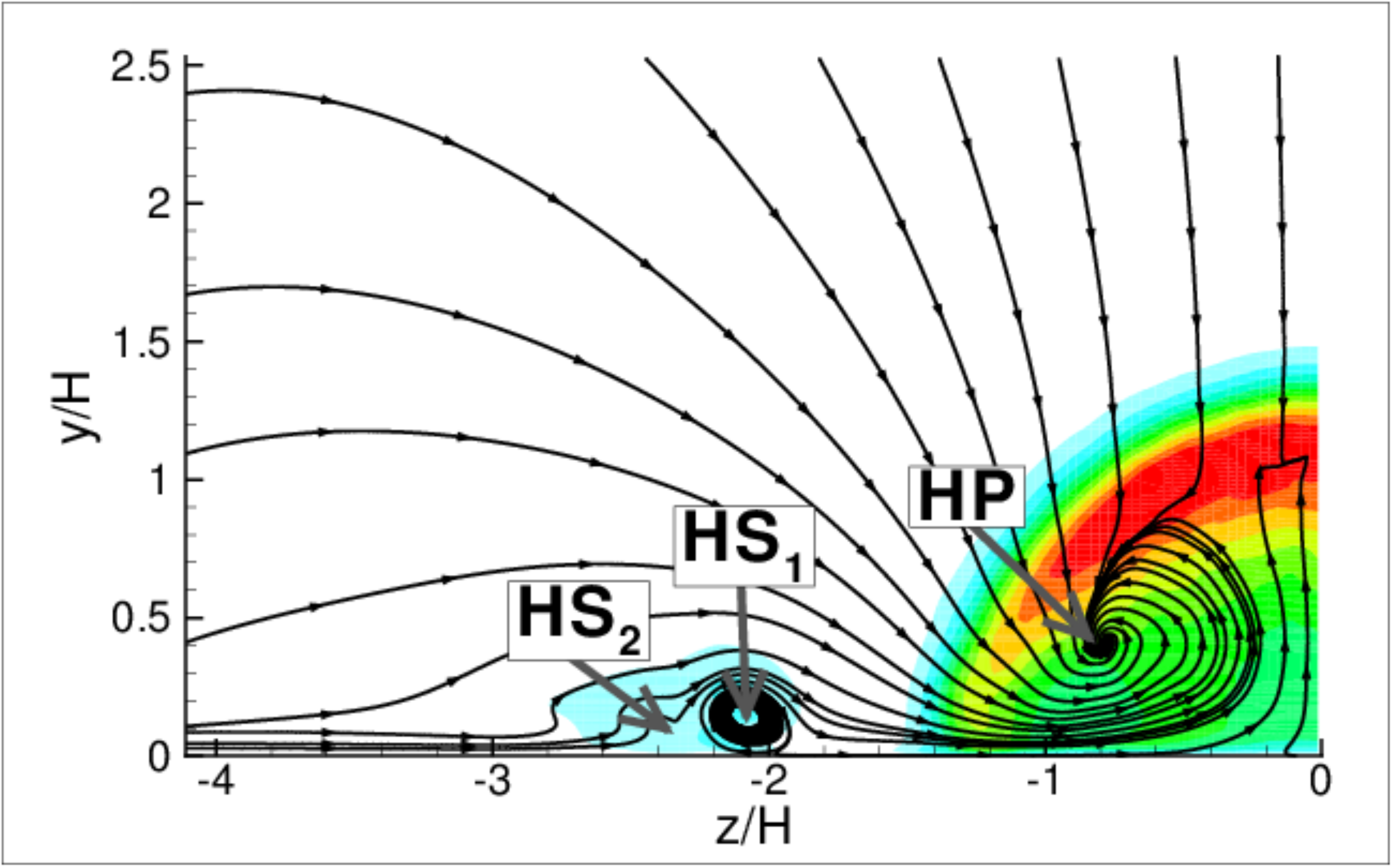}
\includegraphics*[width=0.32\columnwidth,keepaspectratio]{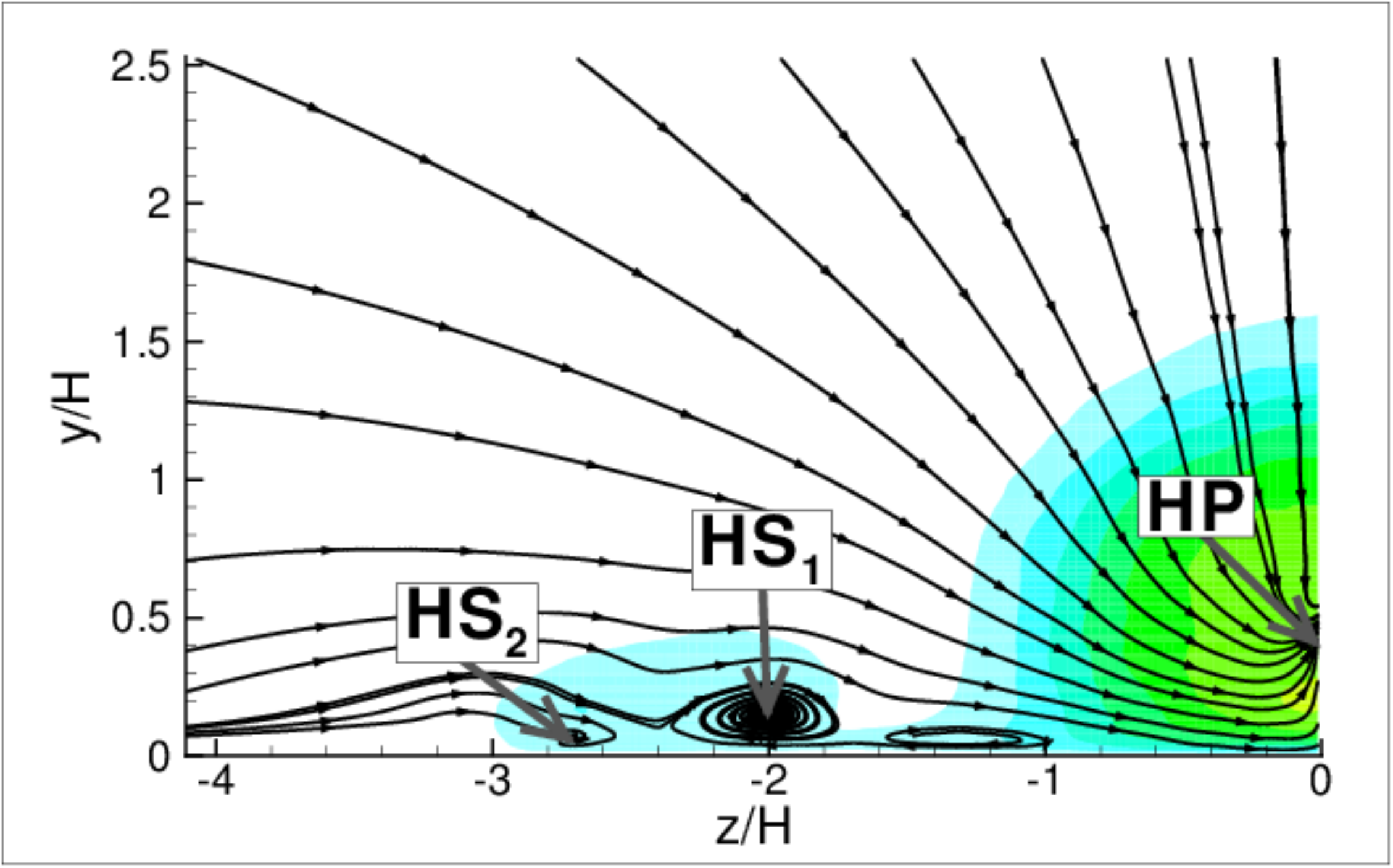}
\includegraphics*[width=0.32\columnwidth,keepaspectratio]{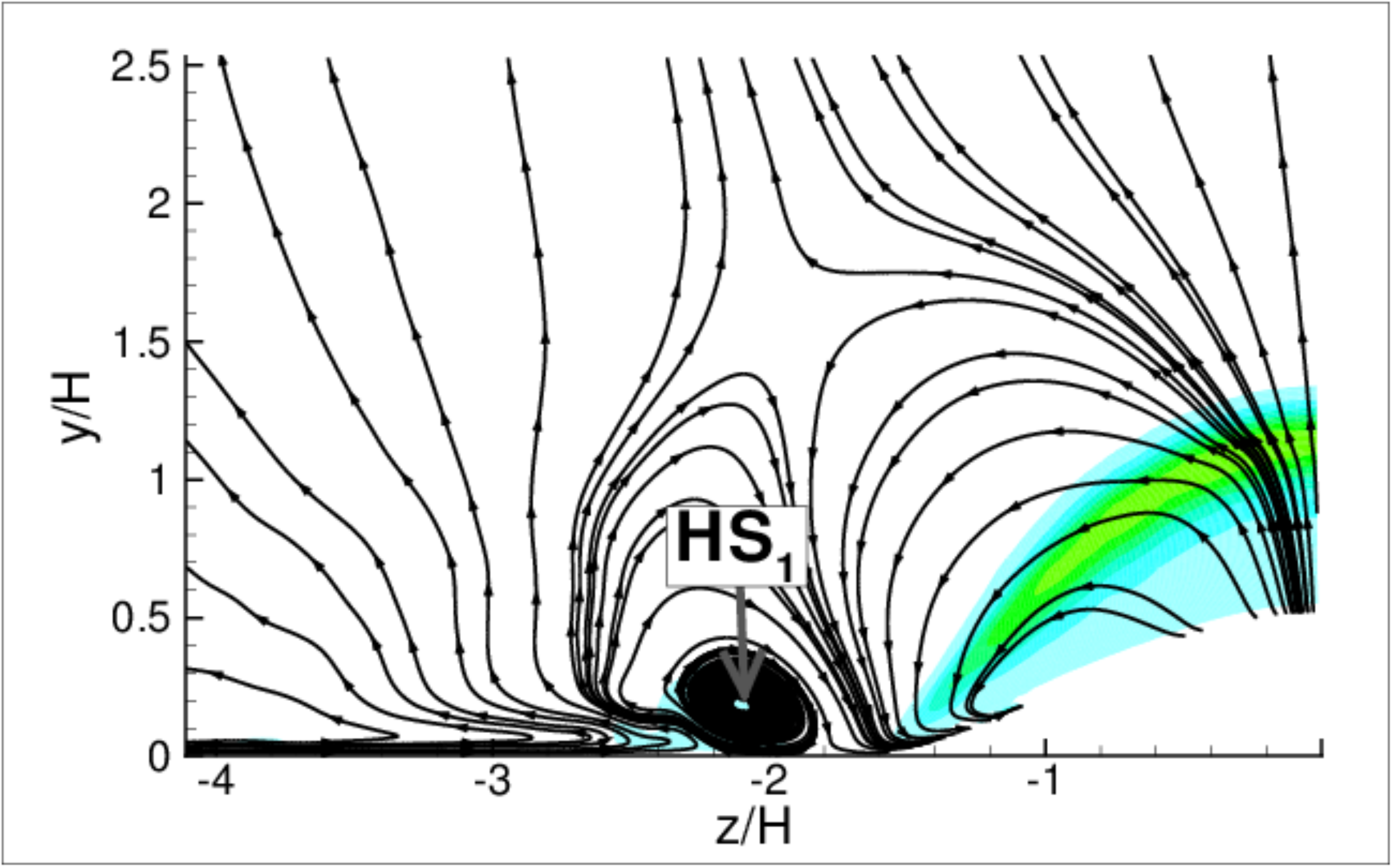}
\includegraphics*[width=0.32\columnwidth,keepaspectratio]{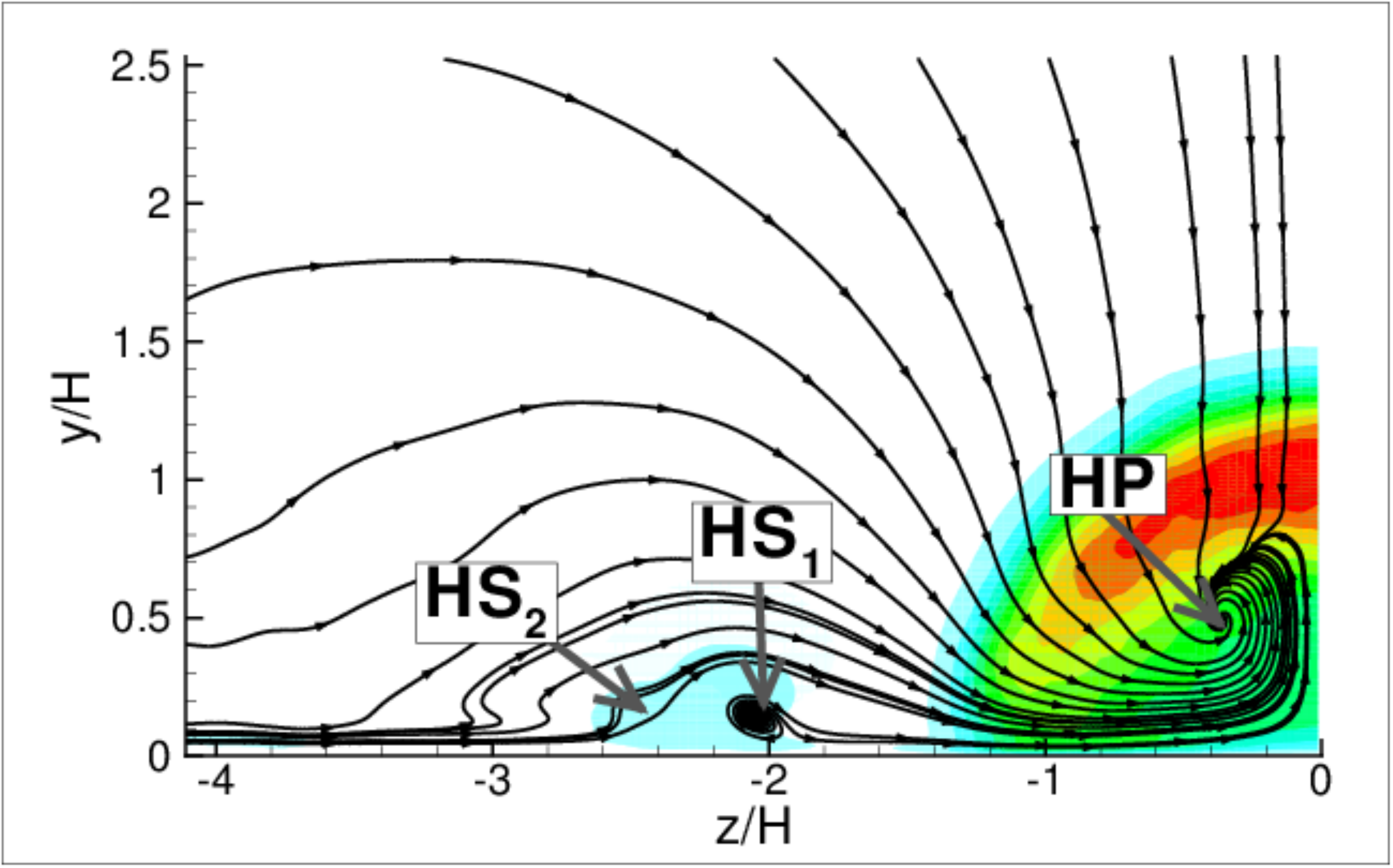}
\includegraphics*[width=0.32\columnwidth,keepaspectratio]{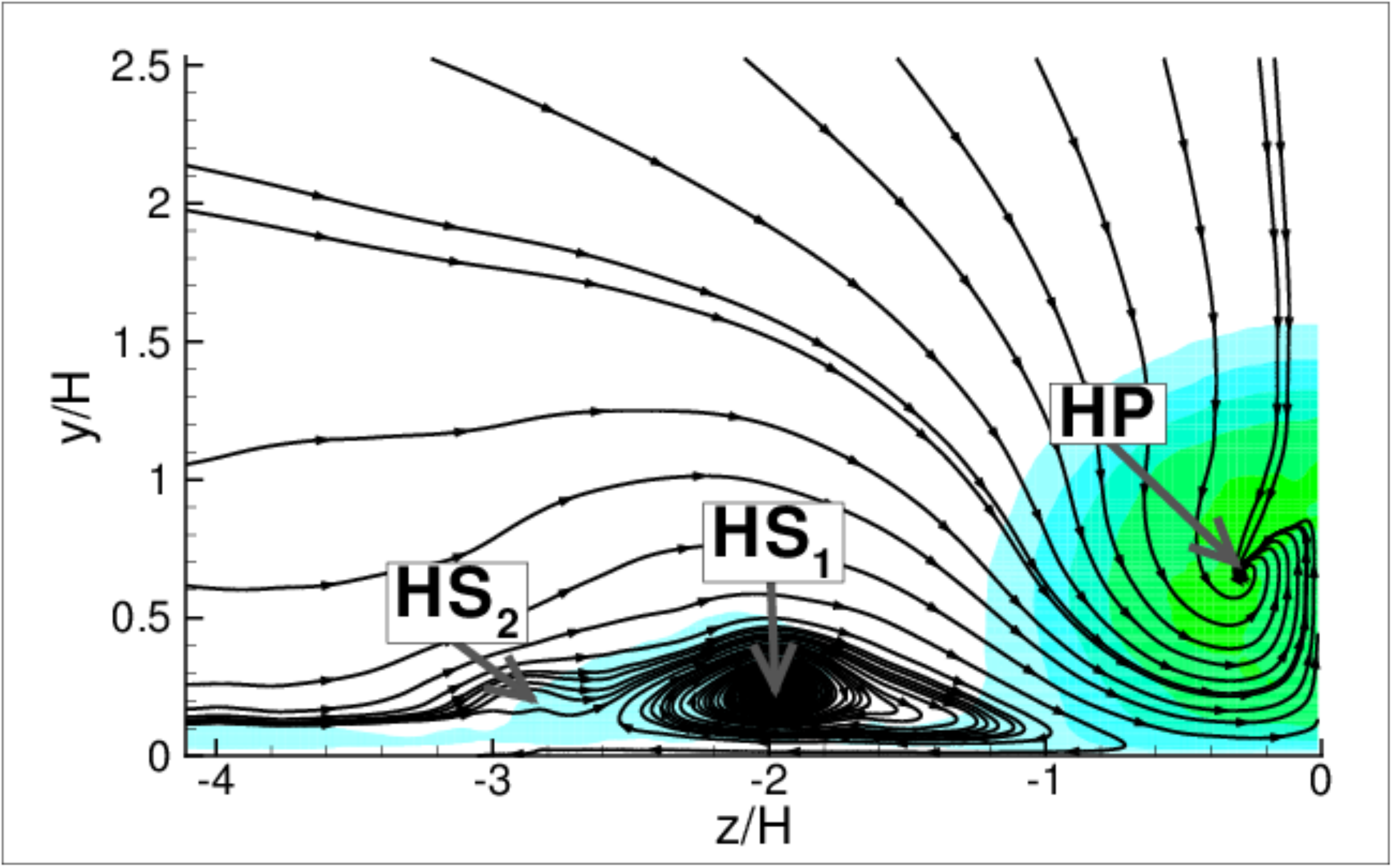}

\end{center}
\caption{Secondary motions. Top, S1. Middle, S2. Bottom, S3. 
Left column, $x/H=1$. Middle column, $x/H=2$. Right column, $x/H=5$.
Color represents turbulent kinetic energy.}
\label{fig:secondary}
\end{figure}

In the wake of the hill, the boundary layer recovers with a combination of 
streamwise acceleration (Fig. \ref{fig:ut_comp}) 
and transverse (secondary) circulation. The secondary
motion is relatively weak with a peak velocity around 10-15\% of $U_{ref}$
and it originates from the realignment of the vorticity generated upstream of
the hill (horseshoe vortex) and additional vorticity shed from the surface of the 
hill. The vorticity generation at the wall and its subsequent 
re-orientation is discussed in \S \ref{sec:vortflux}. 
Fig. \ref{fig:secondary} shows the streamwise evolution of the secondary 
motion at three locations in the near wake of the hill, namely $x/H=1$, 2 and 5.

At $x/H=1$, the legs of the horseshoe vortex are clearly visible around $|z/H|\sim2$.
It is interesting to note that in Simulation S1 two vortices 
($|z/H|\sim2$ and $|z/H|\sim2.8$) are present 
with a counter-rotating region in between. On 
the other hand, in Simulations S2 and S3 only one such vortex 
is visible. Further downstream
for these two simulations there is a trace of  a second vortex (labeled $HS_2$)
although much weaker than in Simulation S1. 

At $x/H=2$, a secondary vortex labelled HP
 appears in all three simulations at $|z/H|<1$. 
Further downstream, at $x/H=5$, for Simulations
S1 and S3 this vortex remains present, with the eye slightly displaced upwards. 
In Simulation S2, this secondary vortex seems to have collapsed in the midplane. 

\begin{figure}
\begin{center}
\includegraphics*[width=\columnwidth,keepaspectratio]{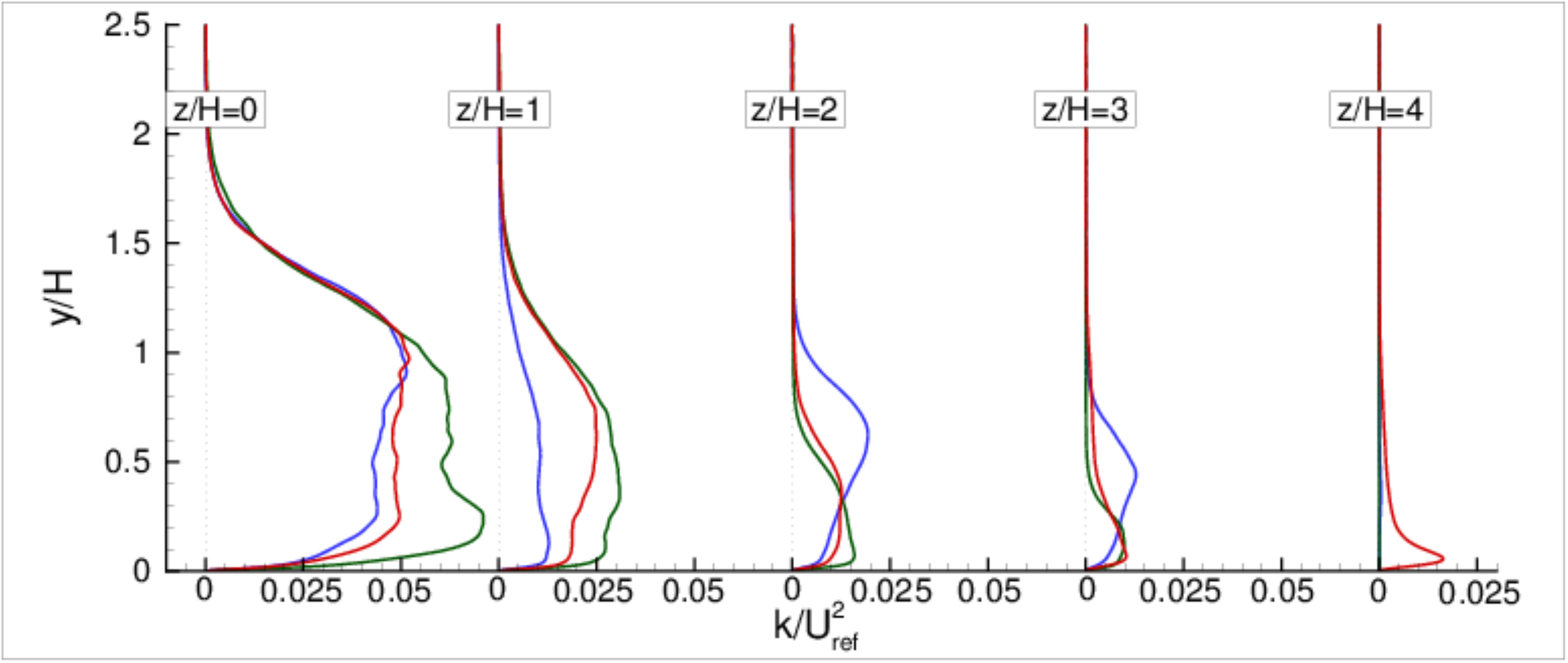}
\end{center}
\caption{Profiles of turbulent kinetic energy at $x/H=5$, 
at various spanwise locations: $z/H=$0, 1, 2, 3, 4.
Line colors defined in Table \ref{tab:1}.}
\label{fig:k_compx5}
\end{figure}

Contours of turbulent kinetic energy $k$ are also included in Fig. \ref{fig:secondary}.
The development of the turbulent kinetic energy in the wake occurs at a later 
streamwise position in Simulation S1 compared to Simulations S2 and S3. At $x/H=1$
patches of $k$ can be observed in the region $|z/H|<1$ in Simulations S2 and S3, 
but not in Simulation S1 (see also Fig. \ref{fig:k_contours}). 
At $x/H=2$, the peak of $k$ is stronger in Simulations S2 and S3. 
The turbulent kinetic energy is concentrated in the shear layer between the free stream
and the recirculation region, in all three simulations. Additionally, there is a 
patch further outwards, which is related to the recirculation in the
horseshoe vortex. Further downstream,
at $x/H=5$, the decay of $k$ with respect to the previous location is clearly 
visible, a typical phenomenon of wakes. In order to illustrate in a more quantitative
manner the differences between the three simulations at 
this streamwise location ($x/H=5$), 
vertical profiles of $k$ at various spanwise locations 
are shown in Fig. \ref{fig:k_compx5}. 
At the midplane ($z/H=0$), $k$ is largest in Simulation S2, by a factor 
of about 20\%, while Simulations S1 and S3 present similar values. At $z/H=1$,
$k$ of Simulation S1 has decreased significantly compared to Simulation S3, indicating
that the width of the wake is smaller in Simulation S1. Further outwards, 
at $z/H=2$ and 3, $k$ of Simulation S1 presents larger values than Simulations S2 
and S3, an in particular at larger heights. This indicates that the horseshoe vortex
is strongest in Simulation S1 and transition to turbulence happens later. 
Finally, at $z/H=4$, all simulations have recovered the
values specified at the inlet, S1 and S2 are laminar while S3 is turbulent. Therefore, 
at this location the boundary layer is undisturbed by the hill.

\subsection{Vorticity flux\label{sec:vortflux}}

Vorticity production at a solid boundary can be described
in terms of vorticity flux. For three-dimensional flows, the mean vorticity
flux can be defined \cite{panton:84,andreopoulos:96} as 
\begin{equation}
\vec{\sigma}=-\nu({\bf n}\cdot  \nabla \vec{\omega})_w, \label{eq:vortflux}
\end{equation}
where $\vec{\sigma}$ is the mean vorticity-flux vector, 
$\vec{n}$ is the normal
vector to the surface, towards the fluid and $\vec{\omega}$ is the 
mean vorticity vector.

\begin{figure}
\begin{center}
\includegraphics*[width=0.9\columnwidth,keepaspectratio]{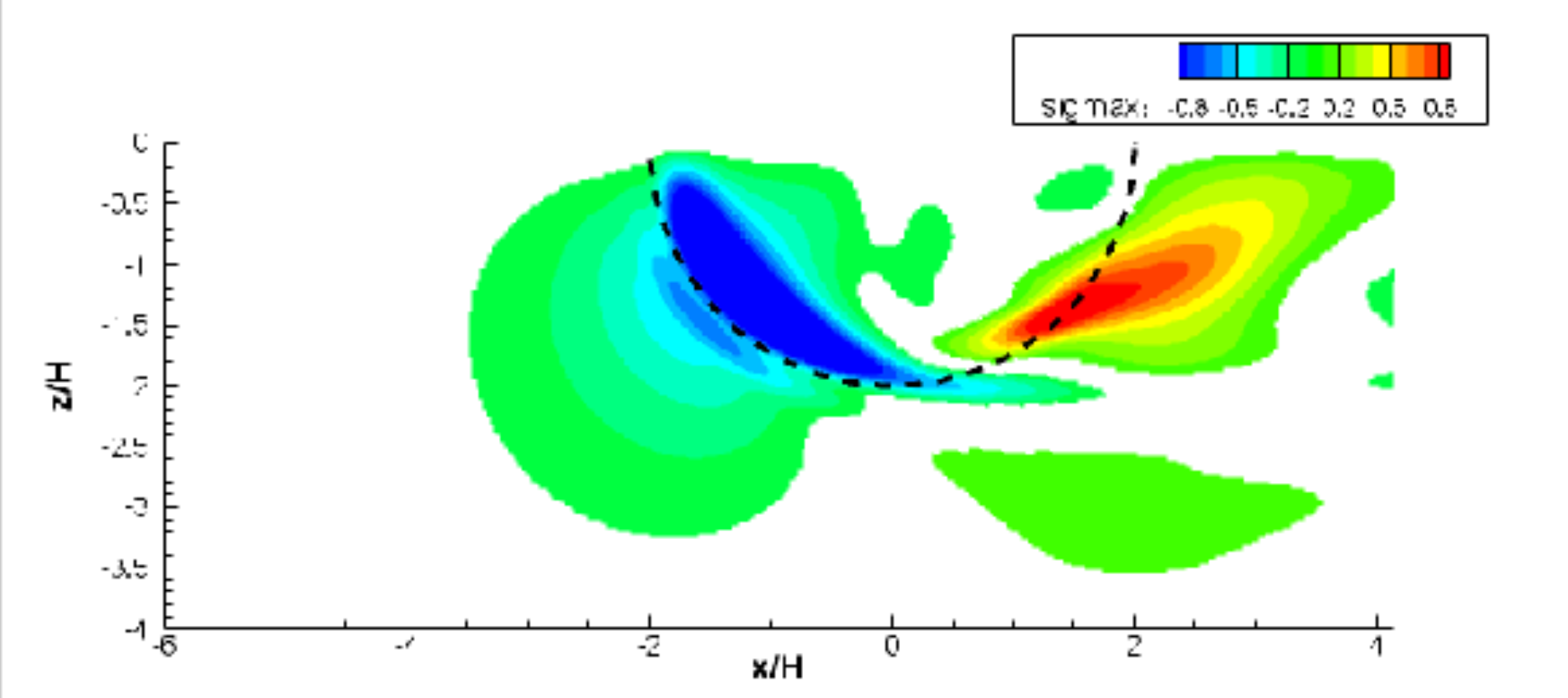}
\includegraphics*[width=0.9\columnwidth,keepaspectratio]{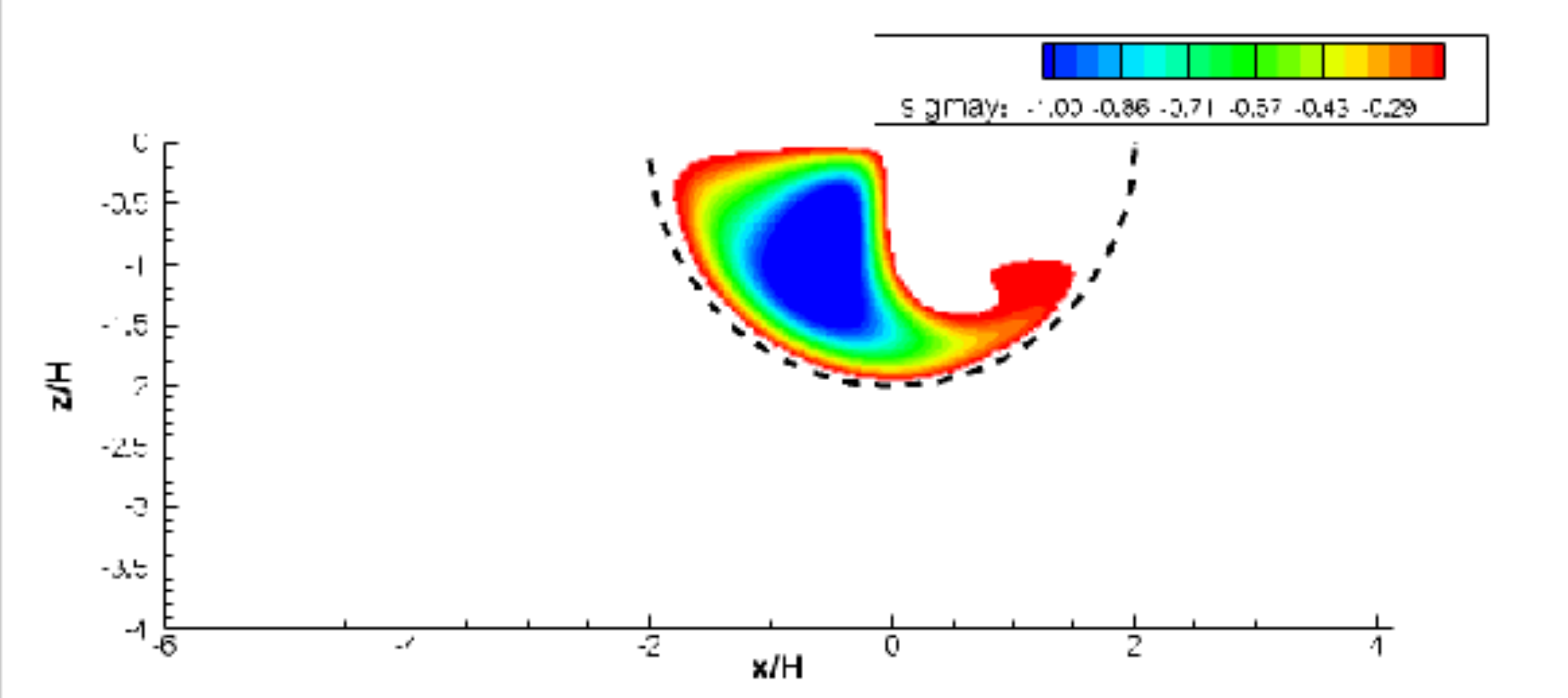}
\includegraphics*[width=0.9\columnwidth,keepaspectratio]{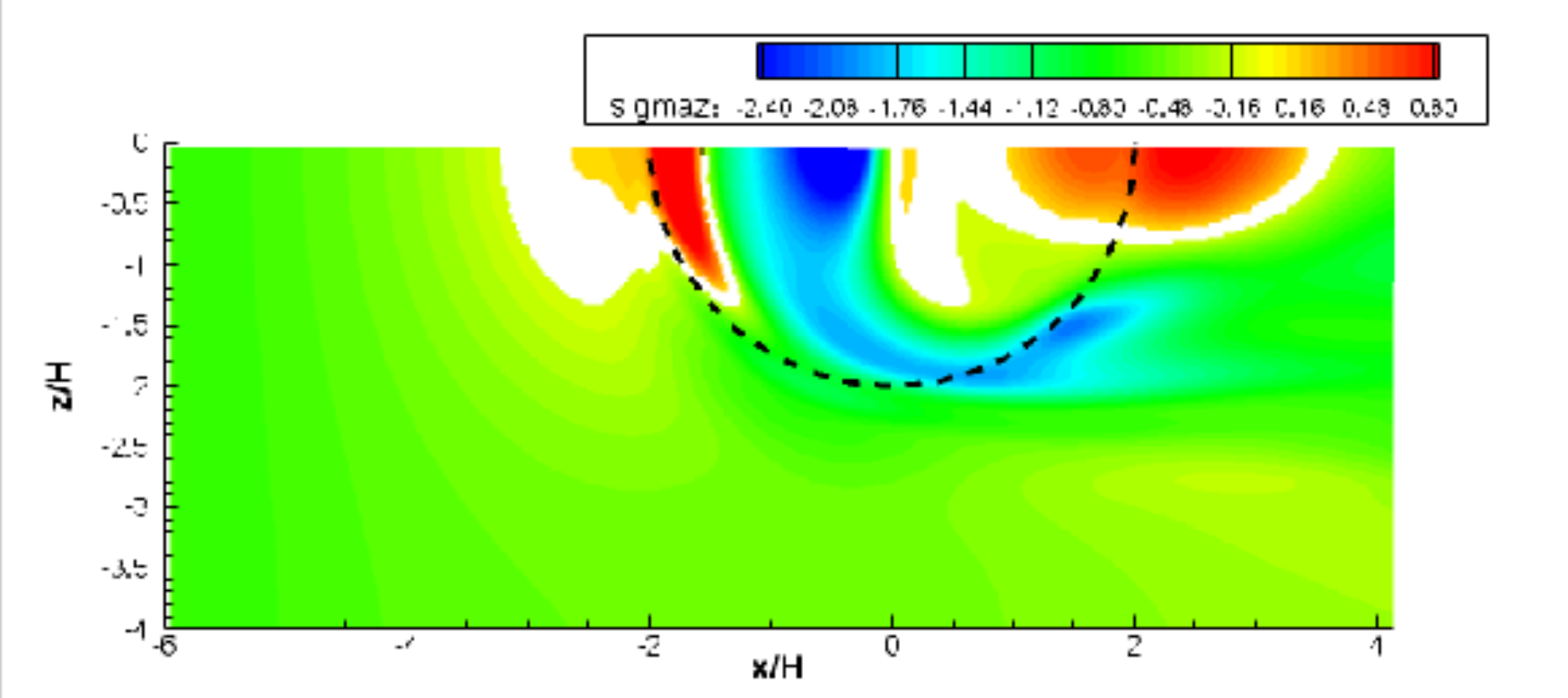}

\end{center}
\caption{Components of the mean vorticity flux at the wall, case S2.
Blanking has been used for $|\sigma|<0.1$
Top, $\sigma_x$. Middle, $\sigma_y$. Bottom, $\sigma_z$.
}
\label{fig:vortflux1b}
\end{figure}

As an illustration, Fig. \ref{fig:vortflux1b} displays the three components
of the mean vorticity flux at the wall from case S2 (the distributions of the other two cases are similar). 
There is production
of spanwise vorticity everywhere in the flow, while the production of 
vertical and streamwise vorticity is concentrated in the hill region. 
Obviously, in the absence of the hill, $\omega_x$ and $\omega_y$ would be zero.
In the vicinity of the hill, $\sigma_z$ (Fig. \ref{fig:vortflux1b}$c$)
reverses sign as a consequence of the reverse flow which occurs both
upstream and downstream of the hill. The negative vorticity flux peaks in the region where the flow strongly accelerates.
Production of $\omega_y$ occurs when the oncoming flow is
deviated sidewards to pass along the left and the right of the hill (only the deviation to the right is illustrated). On the half-domain displayed in Fig. \ref{fig:vortflux1b}$b$,
only negative $\omega_y$ is generated. Finally, production of
$\omega_x$ (Fig. \ref{fig:vortflux1b}$a$) is concentrated towards the side of the
hill: Negative $\omega_x$ is generated upstream of the crest due to the flow 
moving upwards and to the right, while downstream of the crest positive
$\omega_x$  is generated due to the flow moving downwards and to
the left.

\begin{figure}
\begin{center}
\includegraphics*[width=0.9\columnwidth,keepaspectratio]{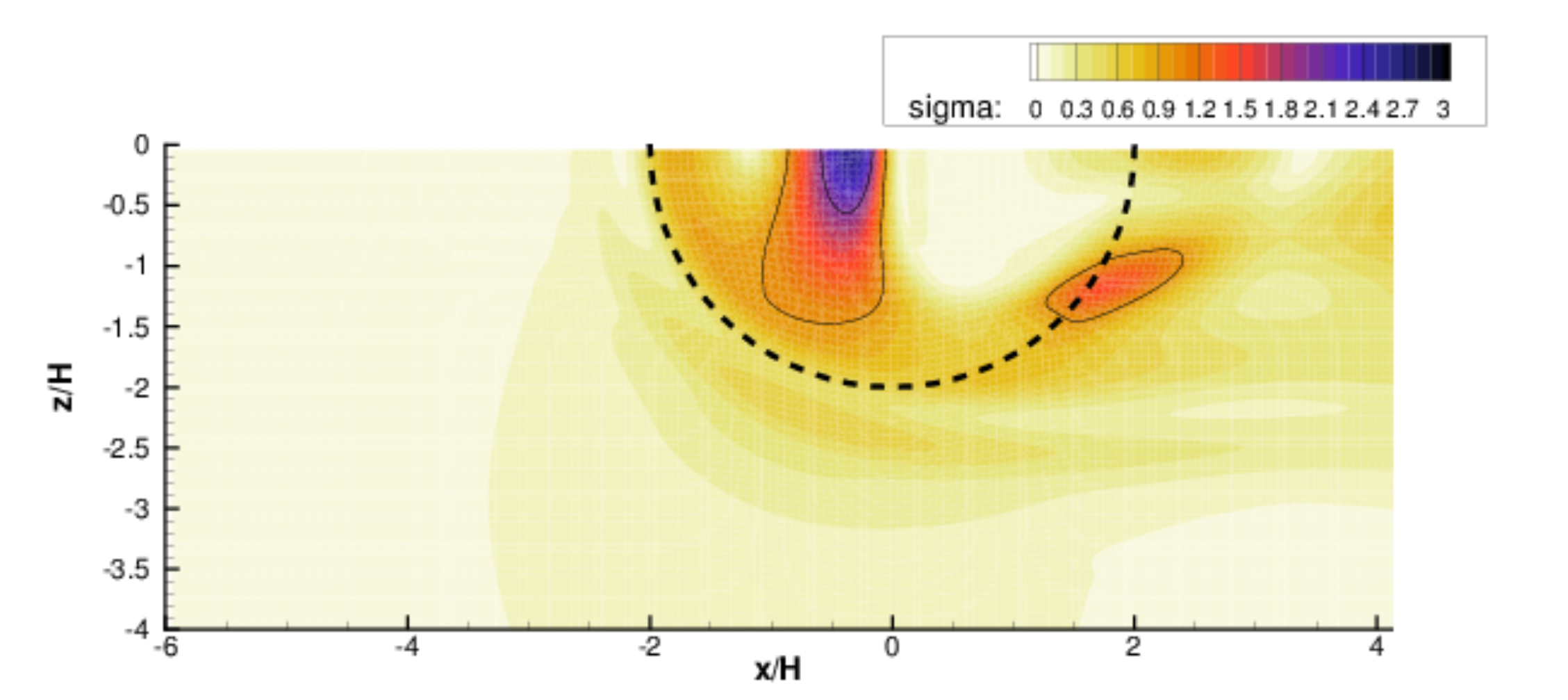}
\includegraphics*[width=0.9\columnwidth,keepaspectratio]{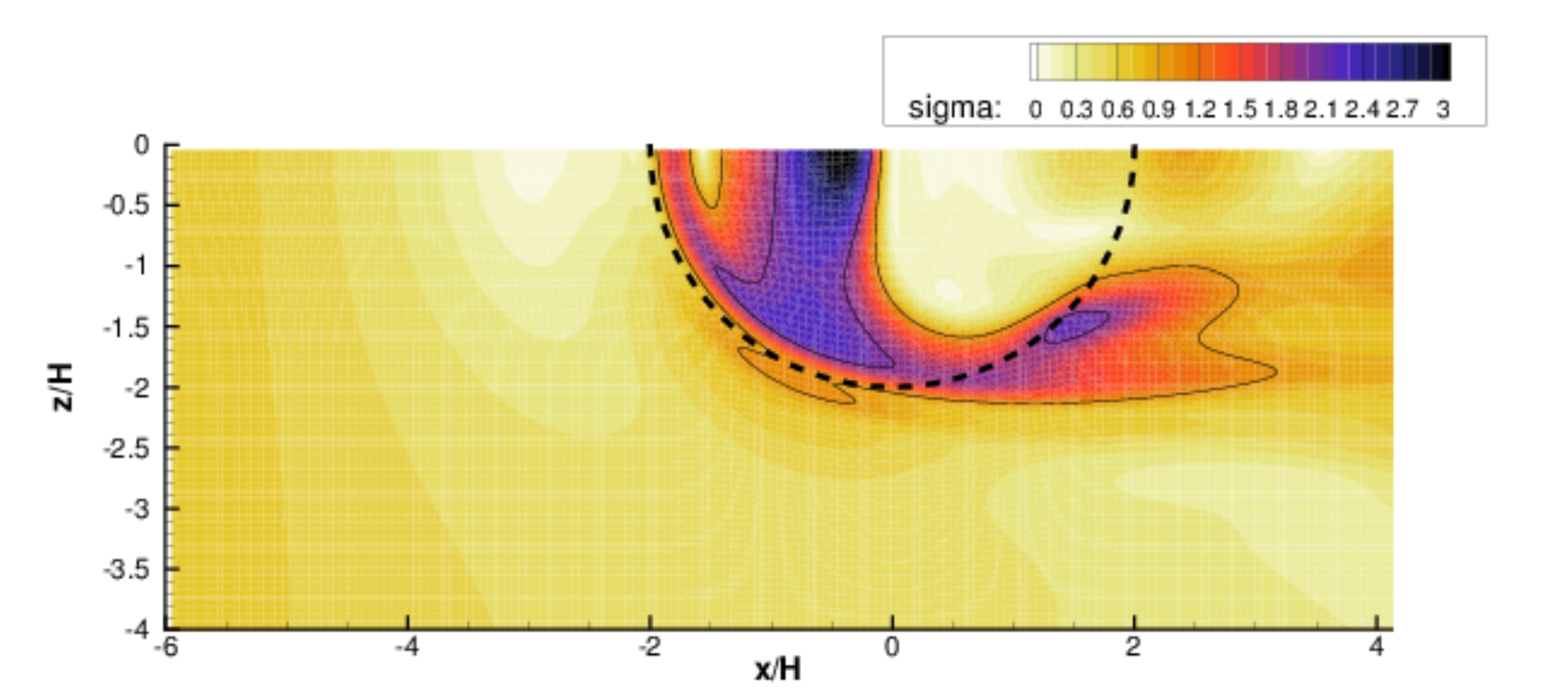}
\includegraphics*[width=0.9\columnwidth,keepaspectratio]{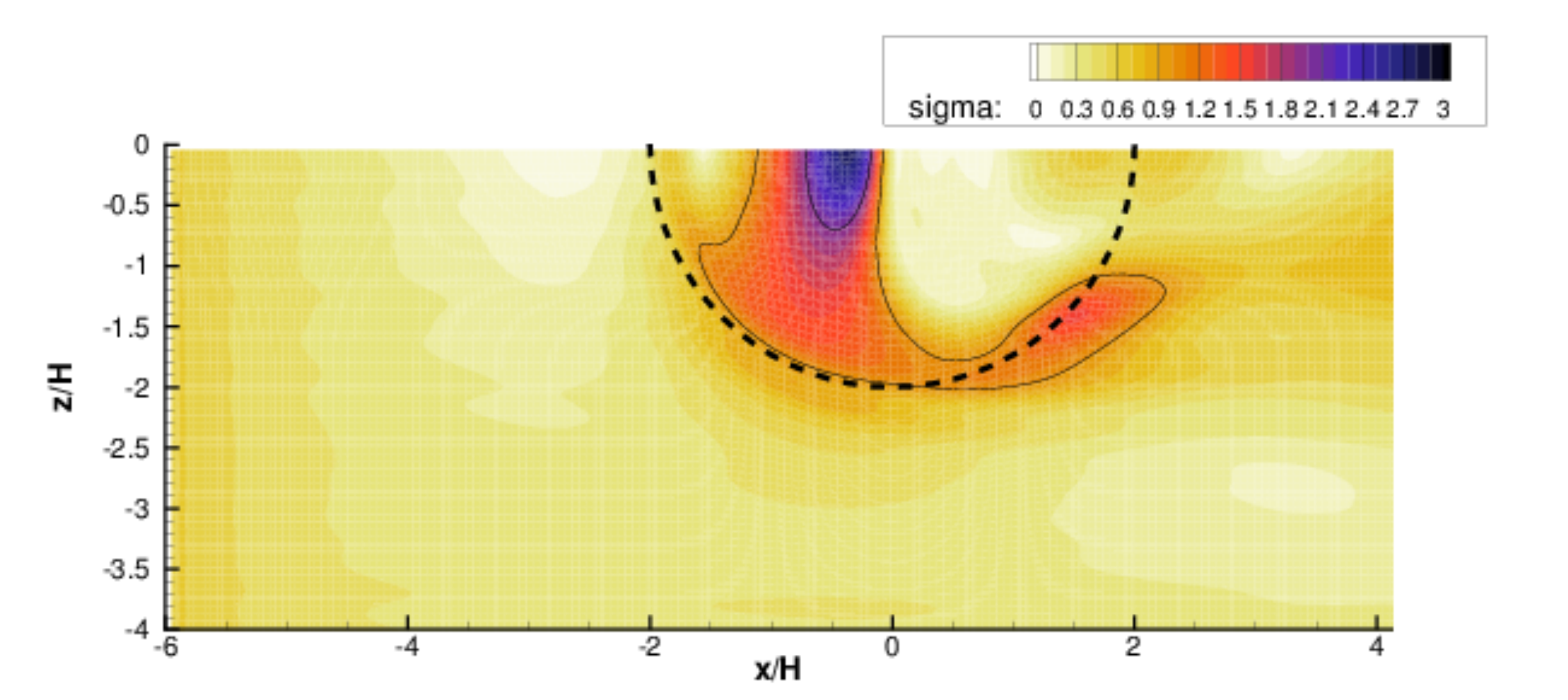}

\end{center}
\caption{Magnitude of the mean vorticity flux at the wall. 
Top, S1. Middle, S2. Bottom, S3.}
\label{fig:vortflux2}
\end{figure}

It is interesting to compare the magnitude of the mean vorticity flux $\sigma=\sqrt{\sigma_i\sigma_i}$ for the three cases. 
This is done in Fig. \ref{fig:vortflux2}. Away from the hill, the values of $\sigma$  are higher in 
cases S2 and S3 compared to S1. This is due to the much lower wall-shear  in case S1 (see Fig. \ref{fig:inlet}).
In the hill region, the maximum values of $\sigma$ are of the same order in all cases ($\sigma \sim 3 U_{ref}^2/H$).
These high values are attained just upstream of the top: 
in the region of strong acceleration of the streamwise velocity (which leads to production of $\omega_z$), 
and towards the sides: where the flow is deviated to the left and to the right (which leads
to production of $\omega_x$ and $\omega_y$). 
It can be seen that, while in all three cases similarly shaped contours are obtained, case S2
has the highest flux, followed by S3 and, finally, S1.
The reason for this is that in S2 more momentum is present below $y/H=1$ (Fig. \ref{fig:inlet}), than in S3, while S3 has more momentum below $y/H=1$ than S1.
The trend (and the argument) is the same as for the pressure coefficient discussed above.
By comparing $\sigma$ in the region of the hill ($\sigma_H$) to its value at the inflow
($\sigma_0$), it is possible to quantify the relative influence of the upstream vorticity and the vorticity generated over the hill.
Due to the low $\sigma_0$ in case S1, $\sigma_H/\sigma_0$ reaches values as high as 50, while in the other two cases 
this ratio is lower; $\sigma_H/\sigma_0\sim8$ in S2 and $\sigma_H/\sigma_0\sim6$ in S3.

\begin{figure}
\begin{center}
\includegraphics*[width=0.6\columnwidth,keepaspectratio]{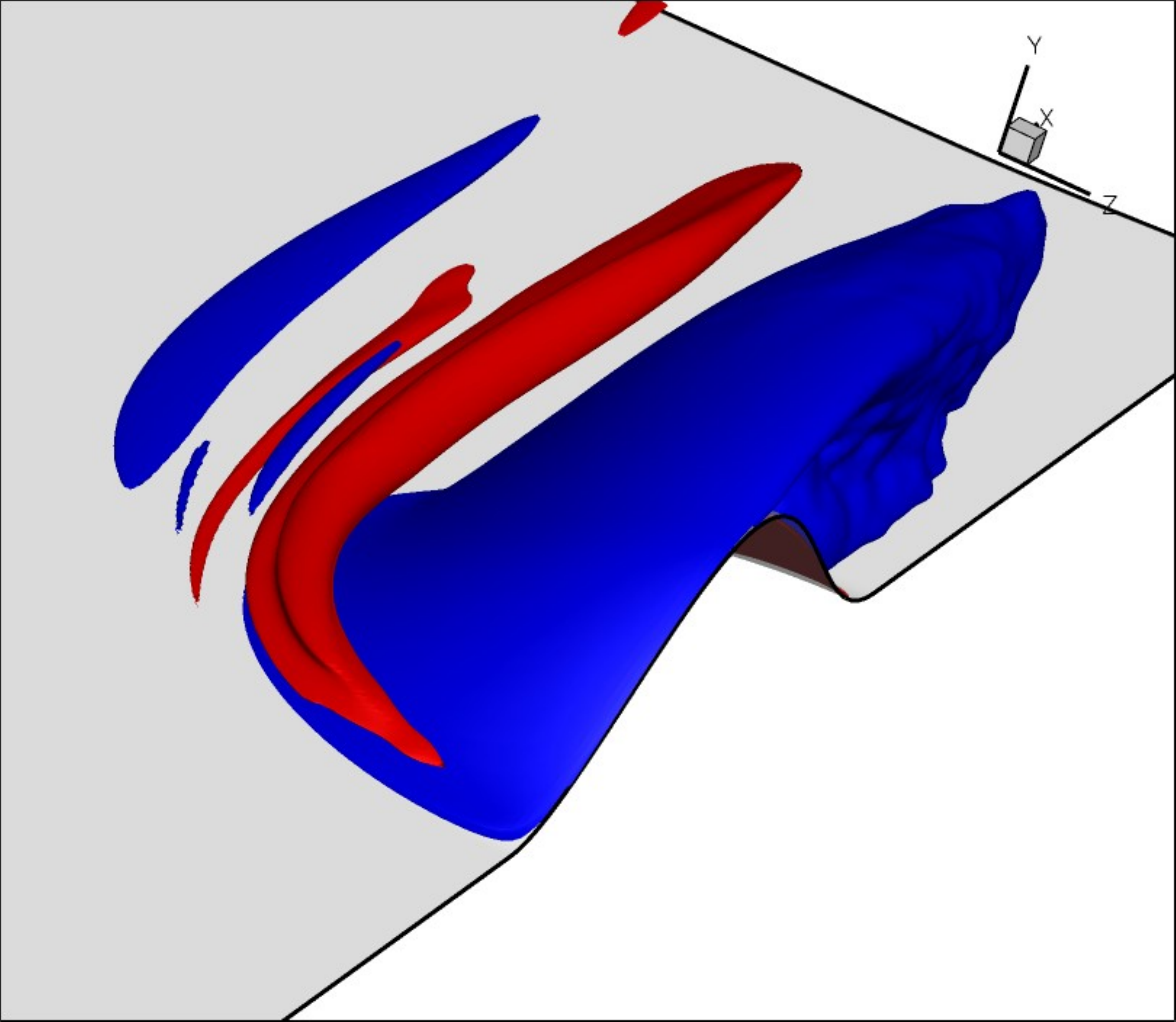}
\end{center}
\caption{Iso-surfaces of mean vertical vorticity from S2. Blue, $\omega_y=-0.5$. Red, $\omega_y=0.5$.}
\label{fig:iso_omey}
\end{figure}

\begin{figure}
\begin{center}
\includegraphics*[width=0.49\columnwidth,keepaspectratio]{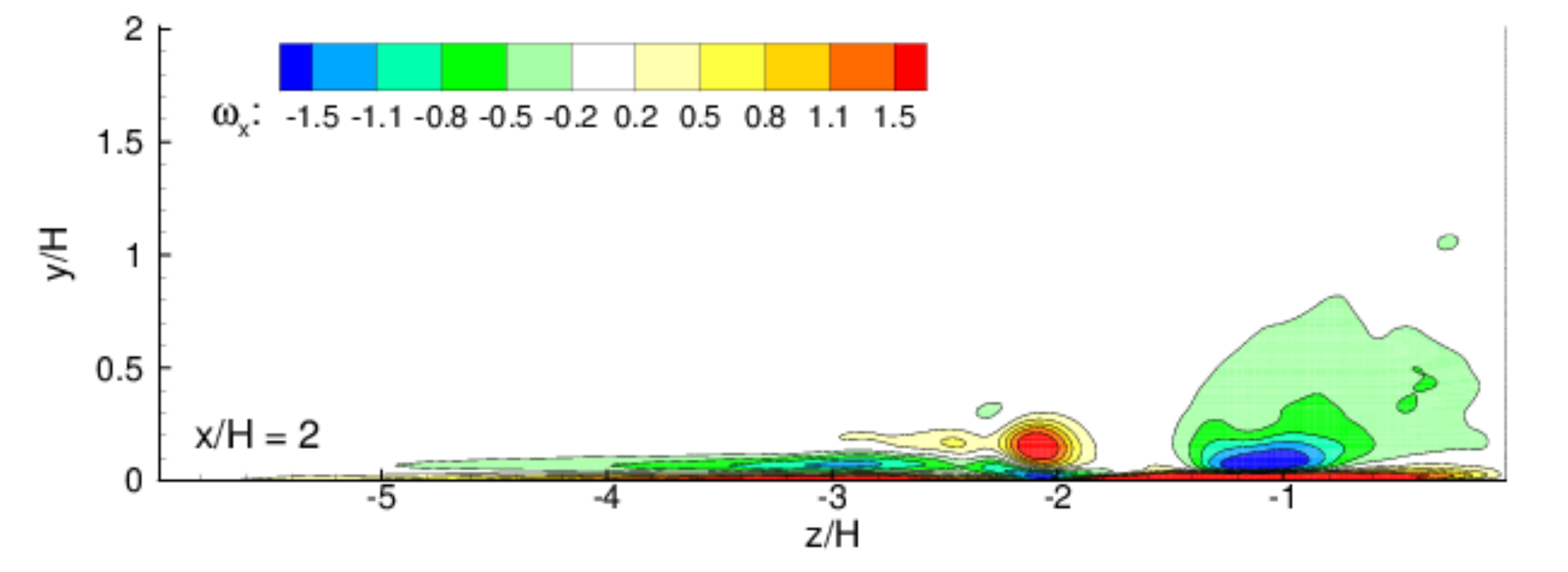}
\includegraphics*[width=0.49\columnwidth,keepaspectratio]{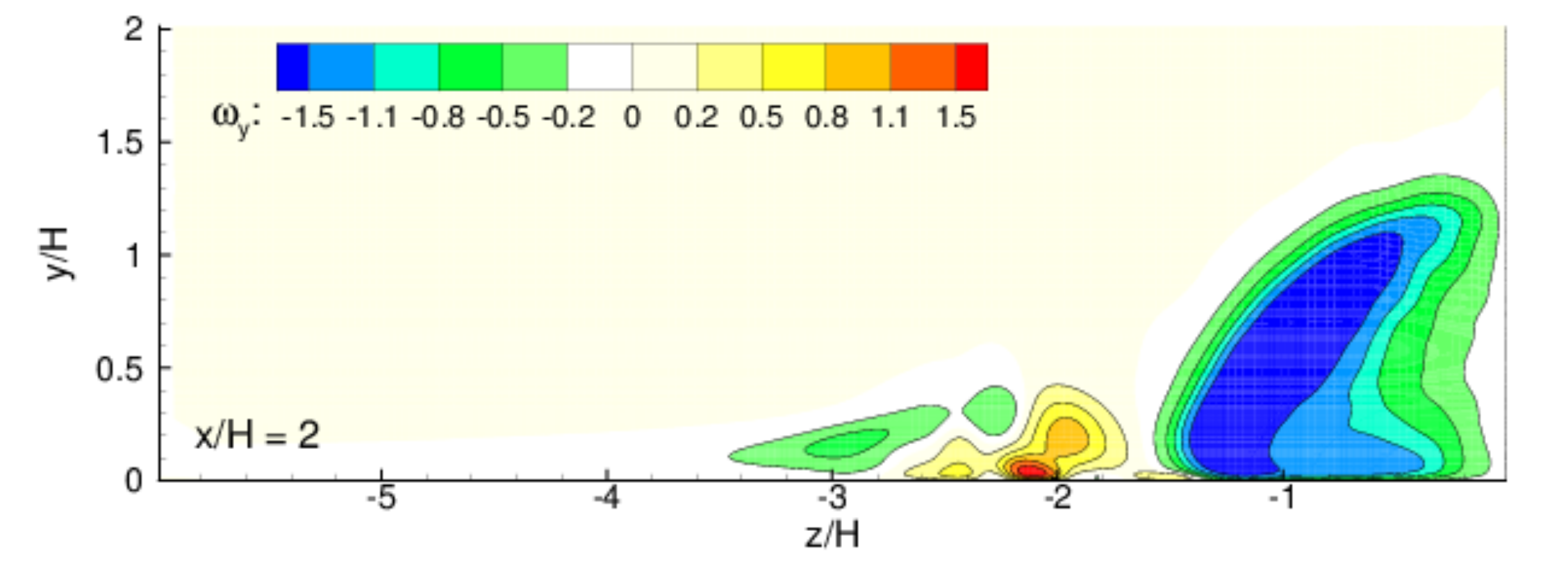}
\includegraphics*[width=0.49\columnwidth,keepaspectratio]{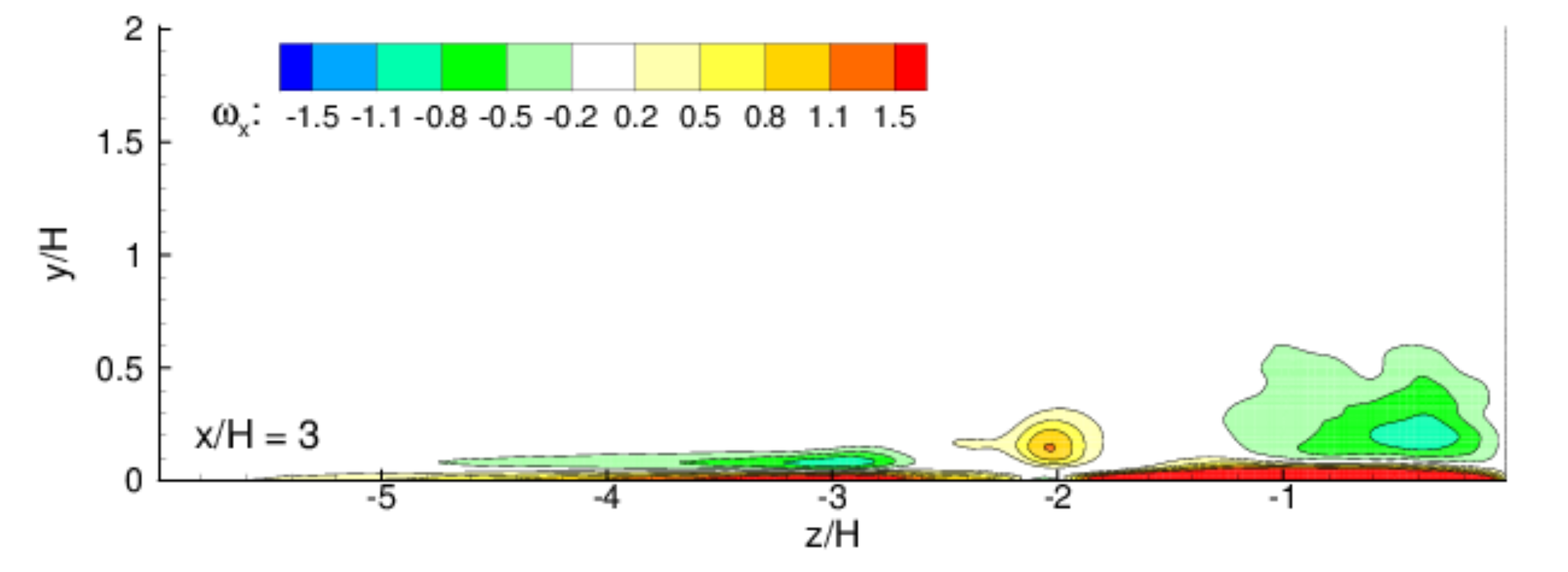}
\includegraphics*[width=0.49\columnwidth,keepaspectratio]{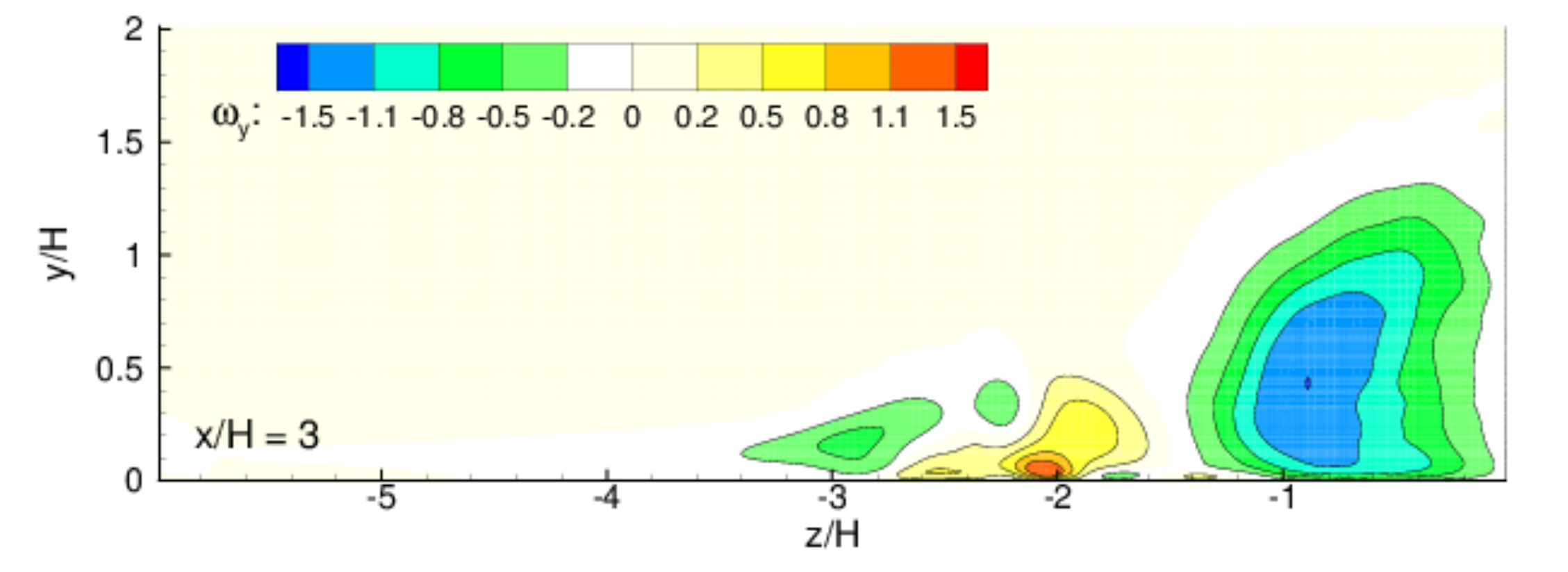}
\includegraphics*[width=0.49\columnwidth,keepaspectratio]{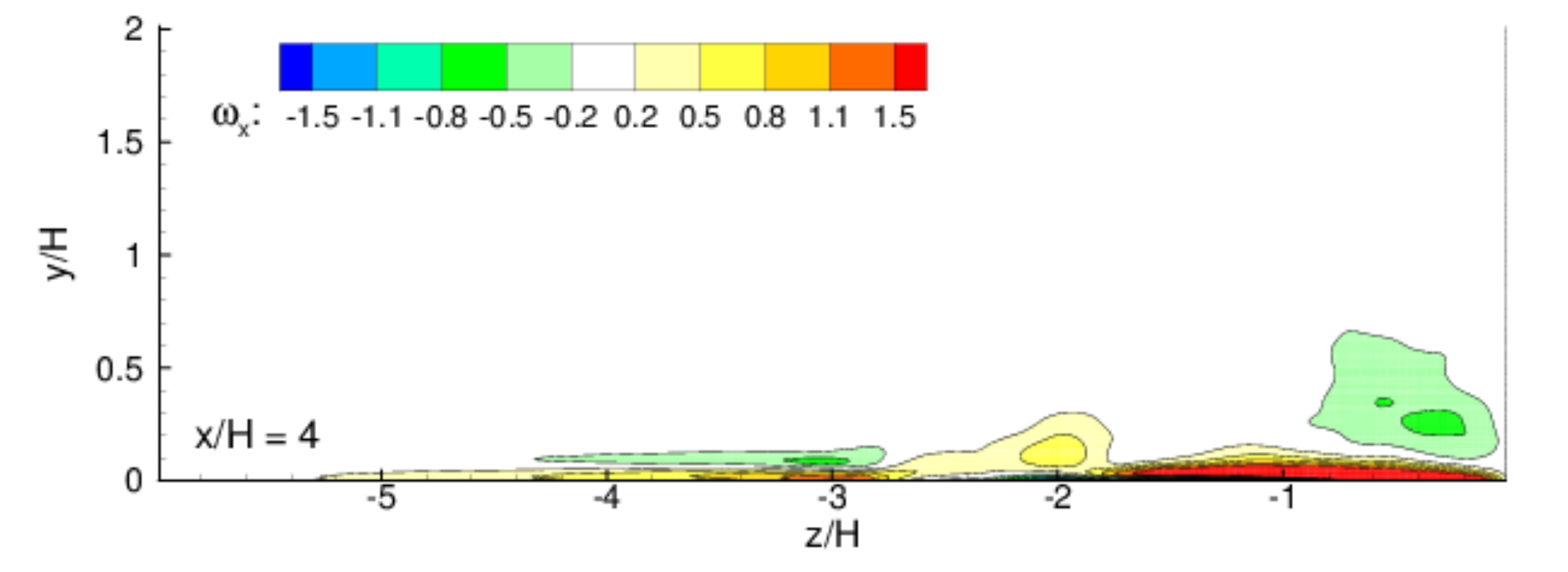}
\includegraphics*[width=0.49\columnwidth,keepaspectratio]{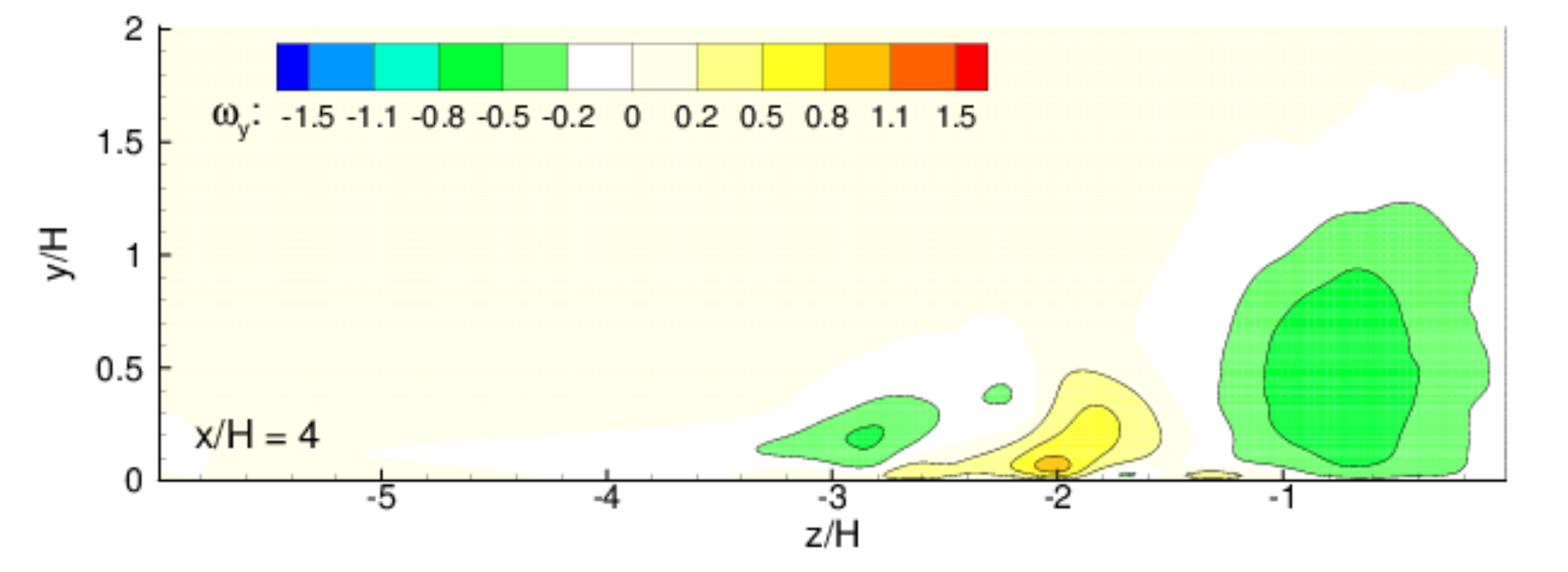}
\includegraphics*[width=0.49\columnwidth,keepaspectratio]{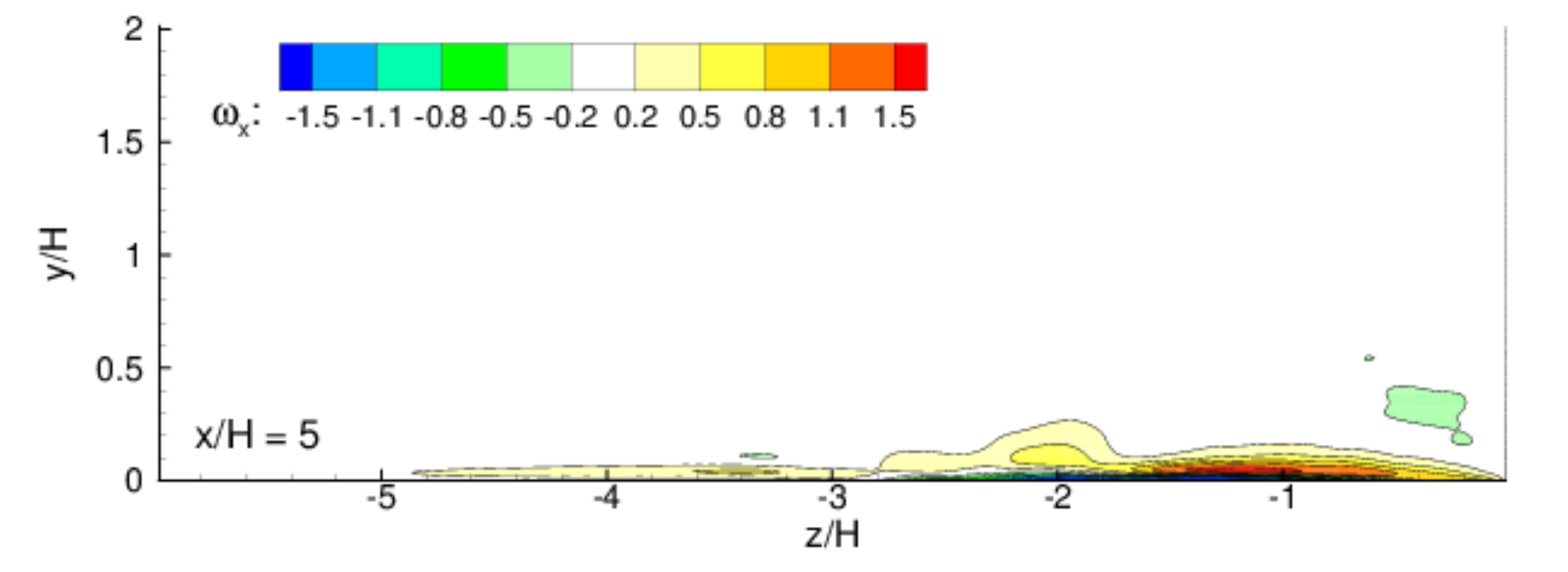}
\includegraphics*[width=0.49\columnwidth,keepaspectratio]{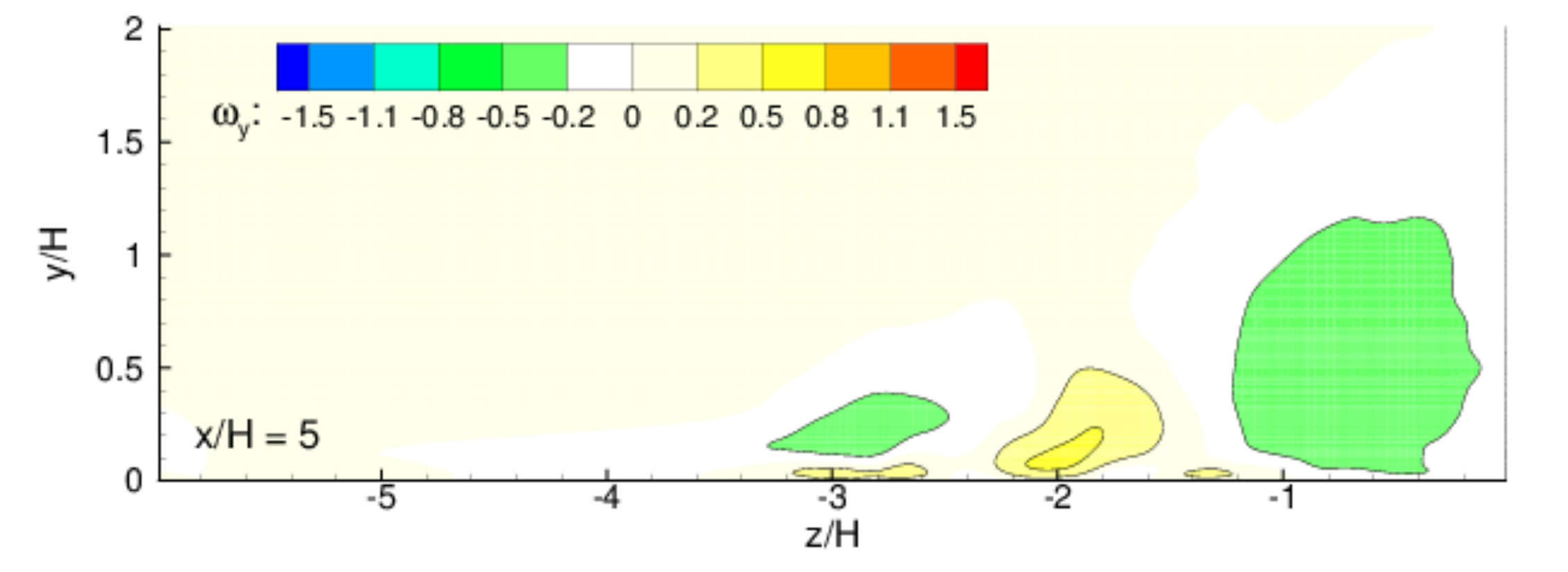}
\includegraphics*[width=0.49\columnwidth,keepaspectratio]{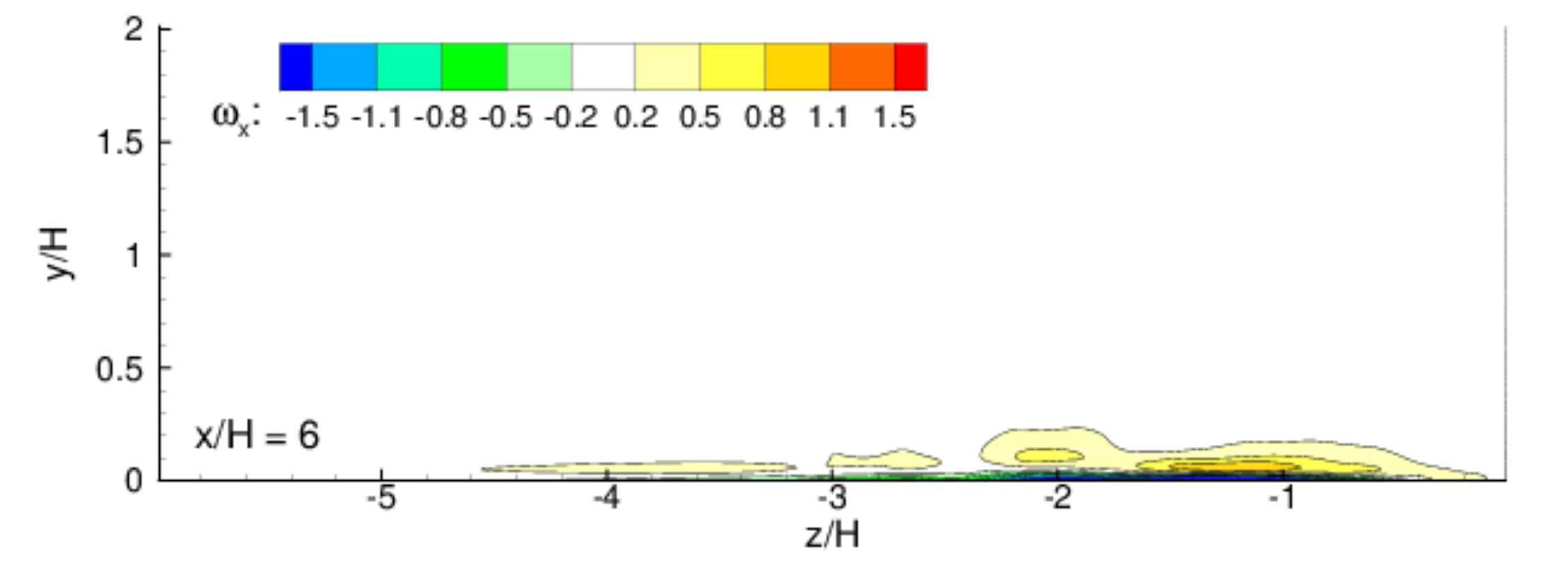}
\includegraphics*[width=0.49\columnwidth,keepaspectratio]{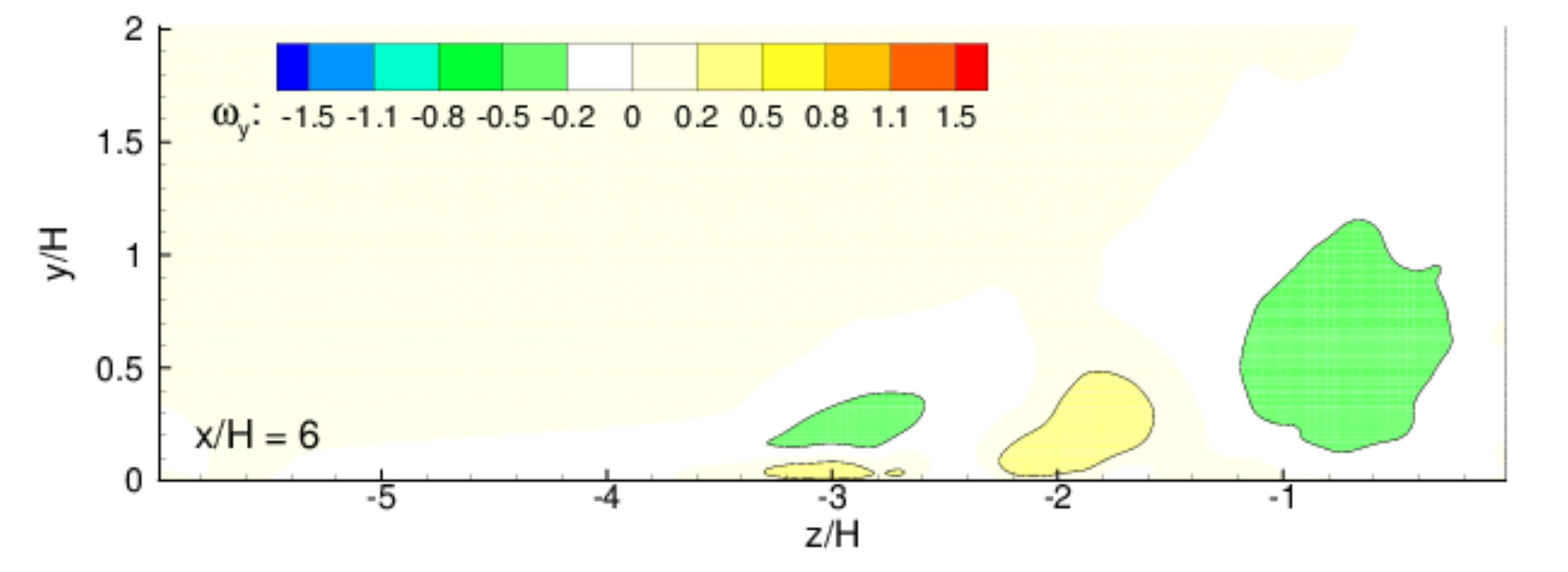}
\end{center}
\caption{Contours of mean streamwise (left) and vertical (right) vorticity from S2.
From top to bottom, $x/H=2$,3,4,5 and 6.}
\label{fig:ome}
\end{figure}

After being produced at the wall, vorticity is subsequently convected and
re-oriented by the mean flow.
This is illustrated in Fig. \ref{fig:iso_omey}, which displays
iso-surfaces of the vertical vorticity for
case S2. Because similar features are observed in the other two cases,
we limit the discussion to S2.
Fig. \ref{fig:vortflux1b}$b$ shows that, for the region
considered, only negative $\omega_y$ is produced
at the wall. The blue iso-surface $\omega_y=-0.5 U_{ref}/H$ originates exactly in
the region of production and is then convected
into the wake. The red iso-surface $\omega_y=0.5 U_{ref}/H$ is, however, not
produced at the wall. Instead, this region
corresponds to the horseshoe vortex and is formed through re-orientation 
of spanwise
vorticity from the incoming boundary layer. Fig. \ref{fig:ome} displays the
values of $\omega_x$ and
$\omega_y$ in the wake of the hill. These two components, which are generated 
as the flow passes over and around the hill, can be seen to be gradually dissipated 
in the downstream direction. 
The vanishing non-spanwise mean vorticity indicates 
that, with increasing distance from the hill, the wake-flow is becoming more and more 
homogeneous in the spanwise direction. 
This figure is related to the secondary motions that were displayed in Fig. \ref{fig:secondary}.
In particular, the patches of $\omega_x$ and $\omega_y$, which are located 
in the region $|z/H| \lesssim 1$, are related to the secondary vortex labelled HP in Fig. \ref{fig:secondary}.
The outer patches are related to the horse-shoe vortex labelled HS1 and HS2 in the same
figure. Therefore, we can conclude that the secondary vortex HP is a direct consequence of the
vorticity production at the surface of the hill while the horse-shoe vortices HS1 and HS2
are only indirectly generated by the hill through re-orientation of vorticity.

\subsection{Instantaneous flow}

\begin{figure}
\begin{center}
\includegraphics*[width=0.6\columnwidth,keepaspectratio]{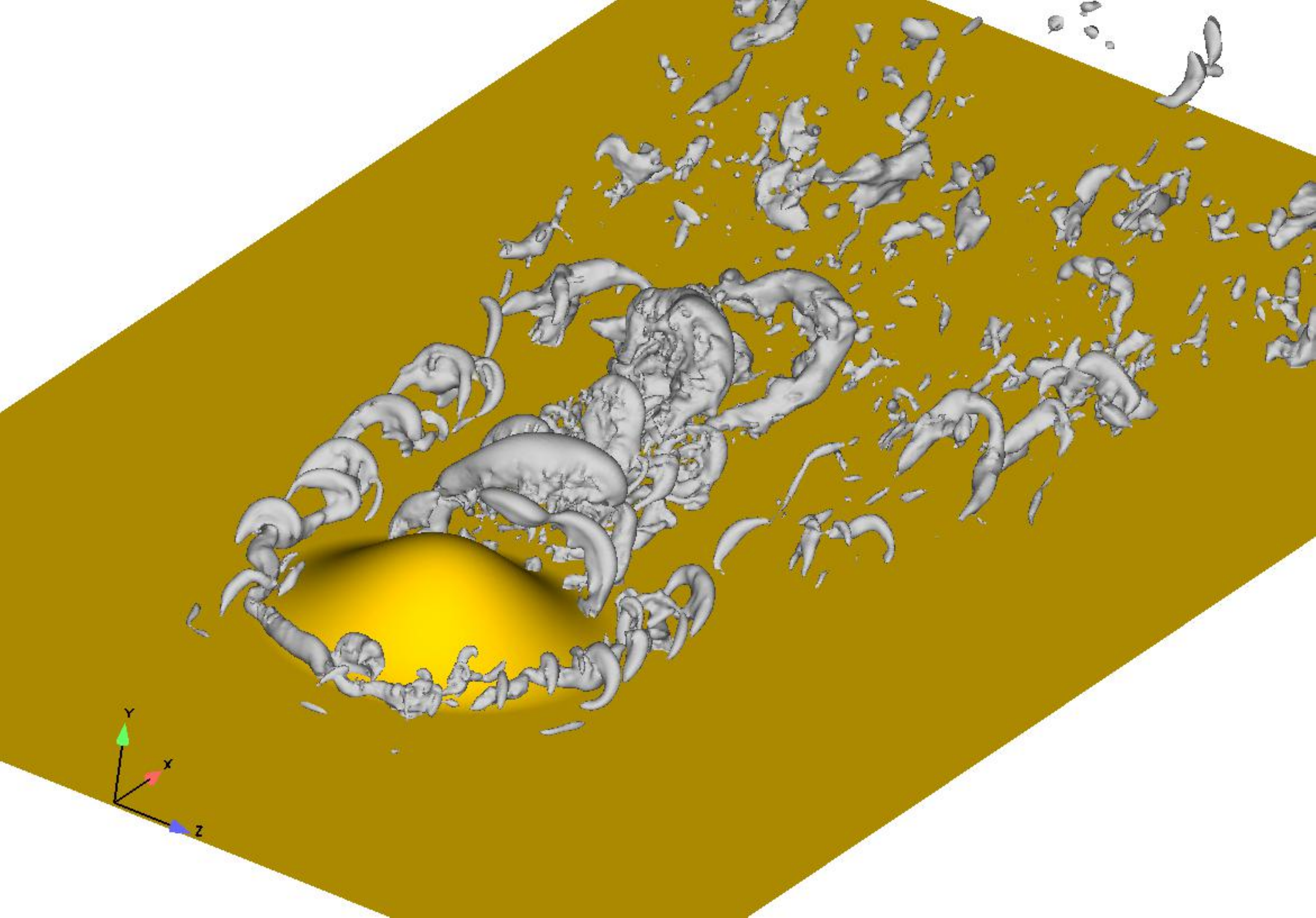}
\includegraphics*[width=0.6\columnwidth,keepaspectratio]{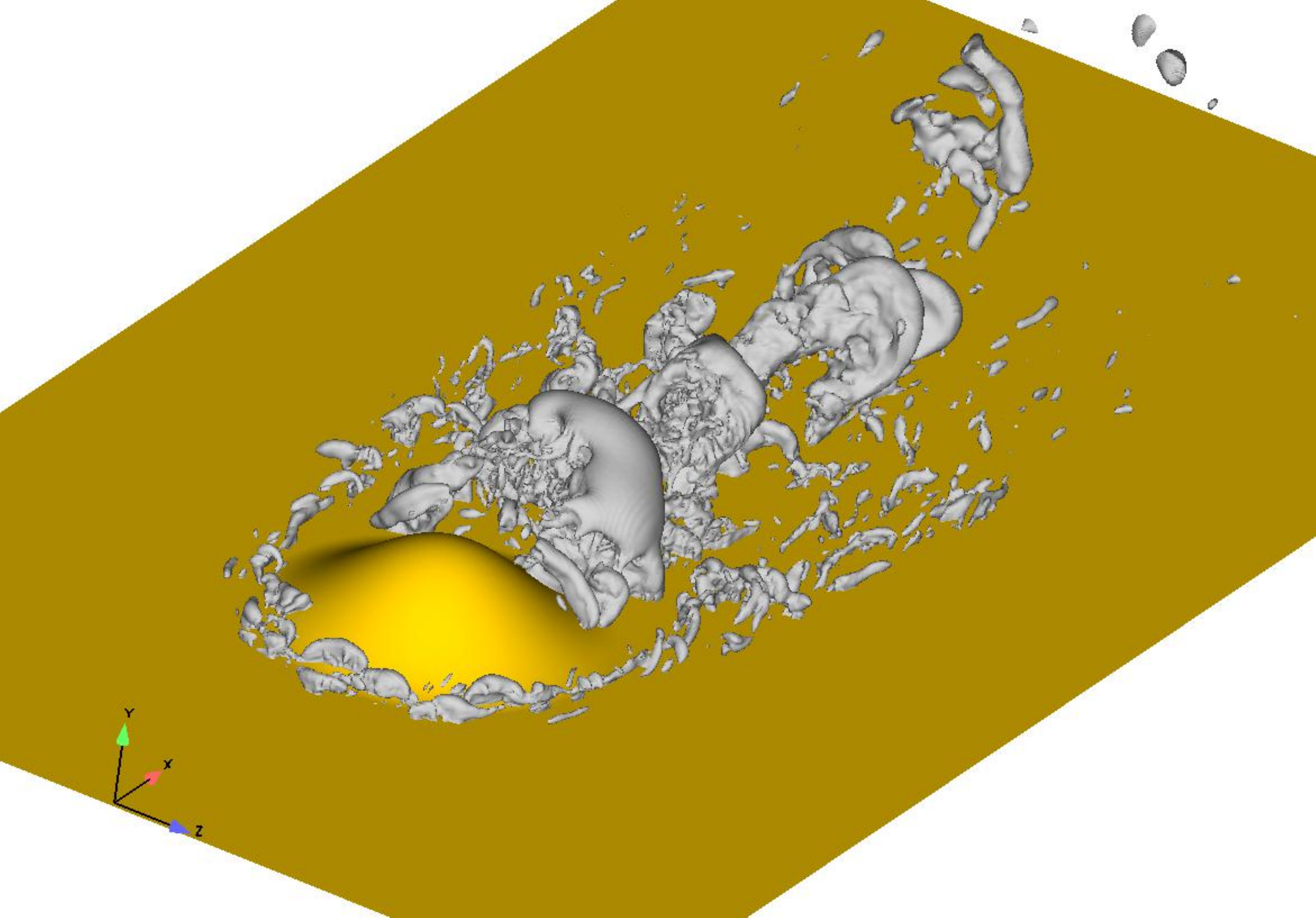}
\includegraphics*[width=0.6\columnwidth,keepaspectratio]{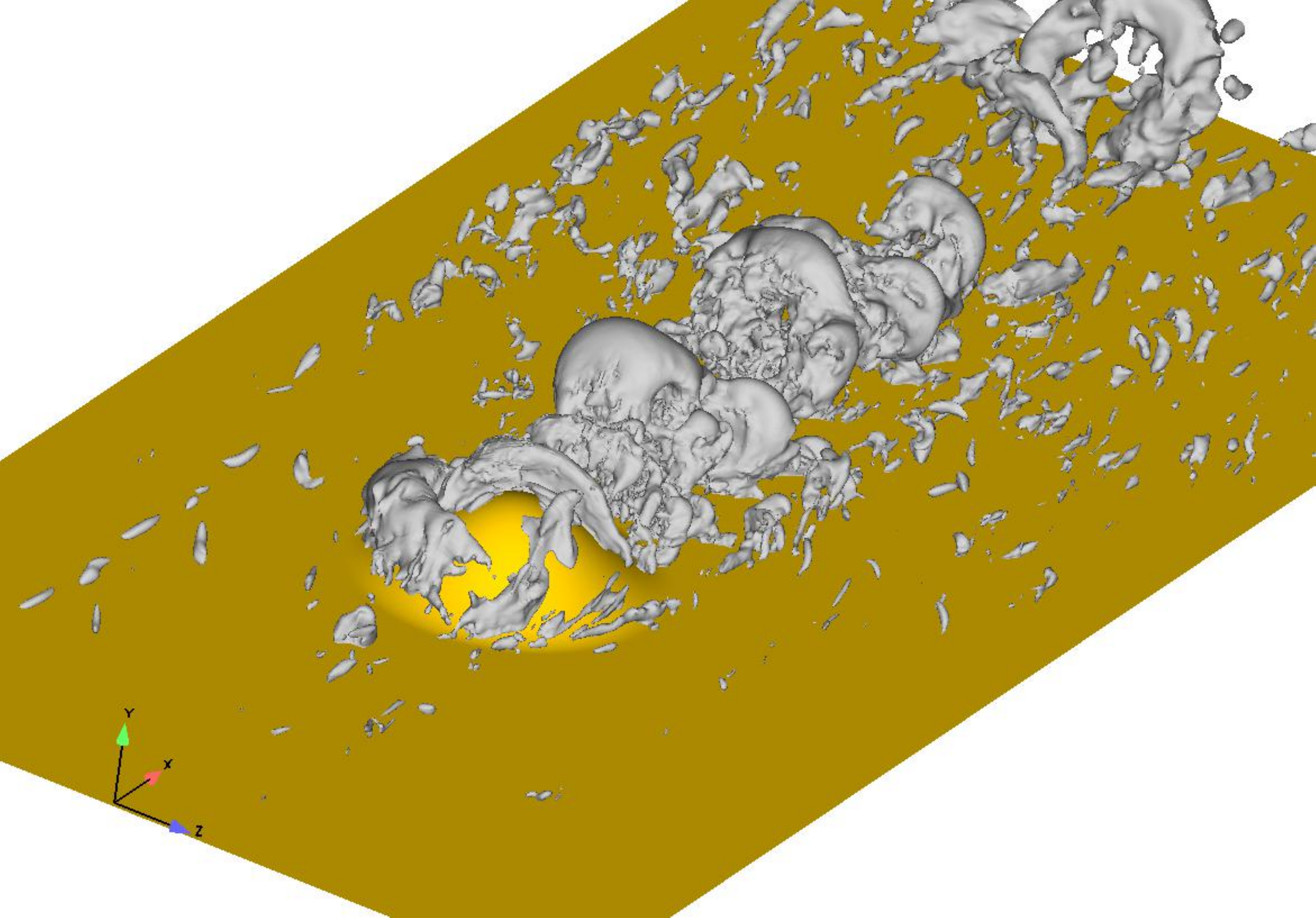}
\end{center}
\caption{Iso-surface of pressure fluctuations. Top, S1- Middle, S2. Bottom, S3.}
\label{fig:struct}
\end{figure}

A visualization of the instantaneous coherent structures of the flow is
displayed in
Fig. \ref{fig:struct}. The figure (and corresponding animations) shows
an iso-surface of pressure fluctuations for the value
$p-\langle p\rangle=-0.02$. This visualization technique has been often used 
in the past \cite{froehlich:05,gv:06}.
Coherent
structures are observed to form in the lee of the hill
and are convected downstream.
Many of them have the shape of a hairpin vortex. Similar structures
have been also observed at high Reynolds number, although in that case,
they were more irregular \cite{gv:09}. It is also well-known that at
lower Reynolds number,
a hemisphere protuberance in a laminar boundary layer
generates a train of very regular hairpin
vortices \cite{acarlar:87}.
It was shown in the cited study that the behaviour
of the wake was quite regular upto $Re_H \sim 3400$. Beyond that
the shedding became irregular. The present investigation ($Re_H=6650$)
lies already in the irregular regime.

An important difference with respect to the high Reynolds number
case is the visibility of
structures originating upstream of the hill
at the reduced Reynolds number considered.
These structures are secondary vortical structures
that appear on top of the horseshoe vortex.
In the present case, these hairpin vortices
are more clearly visible in the cases with laminar inflow,
in particular, in Simulation
S1, see Fig. \ref{fig:struct}(top) and Animation S1. In the middle, a train of large hairpin vortices 
can be observed, while at both sides of the hill a sequence of much 
smaller hairpin vortices can be seen.
In Simulations S2 and S3, (Fig. \ref{fig:struct} middle and bottom, respectively) the large hairpin vortices in the middle can still be
observed, but the smaller vortices at each side of the hill have
become more difficult to identify (Simulation S2 and Animation S2) or almost completely vanished
(Simulation S3 and Animation S3). It appears that the reduced wall-shear in Simulation S1 provides
ideal conditions for the generation of a horseshoe vortex on which secondary
vortical structures are formed that, in the downstream direction, turn into
hairpin vortices. 
Note that the primary horsehoe vortex can only be indirectly detected by
studying iso-surfaces of the pressure fluctuations.
The turbulence, combined with the increased wall-shear stress in Simulation S3
virtually prevent the formation of secondary vortical structures on the 
horseshoe vortex.

%

\section{Conclusions}

In this paper, results of three LES of flow over and around a three-dimensional hill
at moderate Reynolds numbers have been presented.
The Reynolds number is lower than in previous investigations \cite{gv:09}
and this has a significant impact in the wake region. While in the present investigation
the flow is massively separated, leading to a large recirculation region behind the hill, 
at high $Re$ the recirculation region is very shallow \cite{gv:09}. Therefore,
no quantitative comparison with the previous case is reported in this paper.
Two of the simulations have incoming laminar boundary layers of different thicknesses
and the third one has an incoming turbulent boundary layer. It has been shown
that the main features of the flow behind the hill are very similar in
all three simulations. For instance, the 
similar size of the main recirculation bubble behind the hill, 
the presence of a  horseshoe vortex originating inmediately
upstream of the hill, a similar 
wall-topology map and a virtual collapse of the mean streamwise velocity profiles
in the midplane beyond $x/H=5$. In spite of this, there are various differences
which need to be pointed out: 
1) The height of the wake, which increases 
with increasing $Re_{\theta}$ of the inflow profile 
2) The maximum level of kinetic energy in the wake 
varies from 20\% to 30\% depending on the simulation. 
3) The horseshoe vortex is observed to be
 significantly affected by the inflow characteristics. 
4) The main secondary
motion in the central region is found to be quite sensitive
to the actual inflow condition prescribed, which might
have a significant impact on heat and mass transport. 
5) The instantaneous coherent
structures show significant variations as well. 
All these fine details indicate
that, when trying to reproduce physical experiments, 
special care has to be taken
concerning the modelling of the inflow conditions in order to avoid 
observable differences in the region of interest.

\section*{Acknowledgments}
The authors are grateful to the steering committee of the supercomputing 
facilities in Stuttgart for granting computing time on the NEC SX-8. MGV acknowledges the financial support of the German Research Foundation (DFG).

%

 \bibliographystyle{plainnat}
 \bibliography{biblio.bib}

\end{document}